\documentclass[12pt]{article}
\usepackage[utf8]{inputenc}
\usepackage[T1]{fontenc}
\usepackage{amsmath, amssymb}
\usepackage{graphicx}
\usepackage{float}
\usepackage{color}
\usepackage[table,xcdraw]{xcolor}
\usepackage{booktabs}
\usepackage{hyperref}
\usepackage{geometry}
\usepackage{caption}
\usepackage{makecell}
\usepackage{lineno}
\usepackage{adjustbox}
\usepackage{longtable}
\usepackage{placeins}
\usepackage{xr}

\title{Sparsity is All You Need: Rethinking Biological Pathway-Informed Approaches in Deep Learning}

\author{
	\begin{tabular}{c}
		Isabella Caranzano$^{1}$, Corrado Pancotti$^{1}$, Cesare Rollo$^{1}$, Flavio Sartori$^{1}$, \\ 
		Pietro Li\`o$^{2}$, Piero Fariselli$^{1}$, Tiziana Sanavia$^{1}$ \\[0.5em]
		$^{1}$Computational Biomedicine Unit, Department of Medical Sciences, \\
		University of Torino, Torino, Italy \\ 
		$^{2}$Department of Computer Science and Technology,\\ University of Cambridge, Cambridge, UK
	\end{tabular}
}
\date{}

\begin{document}
	
\maketitle

\begin{abstract}
    Biologically-informed neural networks typically leverage pathway annotations to enhance performance in biomedical applications. We hypothesized that the benefits of pathway integration does not arise from its biological relevance, but rather from the sparsity it introduces. \\
    We conducted a comprehensive analysis of all relevant pathway-based neural network models for predictive tasks, critically evaluating each study’s contributions. From this review, we curated a subset of methods for which the source code was publicly available. The comparison of the biologically informed state-of-the-art deep learning models and their randomized counterparts showed that models based on randomized information performed equally well as biologically informed ones across different metrics and datasets. Notably, in 3 out of the 15 analyzed models, the randomized versions even outperformed their biologically informed counterparts. 
    Moreover, pathway-informed models did not show any clear advantage in interpretability, as randomized models were still able to identify relevant disease biomarkers despite lacking explicit pathway information. \\
    Our findings suggest that pathway annotations may be too noisy or inadequately explored by current methods. Therefore, we propose a methodology that can be applied to different domains and can serve as a robust benchmark for systematically comparing novel pathway-informed models against their randomized counterparts. This approach enables researchers to rigorously determine whether observed performance improvements can be attributed to biological insights.

\end{abstract}

\flushbottom
\maketitle

\thispagestyle{empty}

\section*{Background \& Summary}

When dealing with deep learning models, many functions that are efficiently computable through a machine learning approach exhibit what is called “compositional sparsity”, meaning that they can be decomposed into a few simpler functions, each depending on only a small subset of inputs. Deep networks, such as Convolutional Neural Networks (CNNs) and Transformers, align with the compositional structure of many target functions, leading to better generalization since they approximate such functions efficiently without falling victim to the “curse of dimensionality”, i.e. the exponential growth of computational complexity with input dimension \cite{wen2016learning,hastie2015statistical,Poggio:2022,Hoefler:2021,poggio2024compositional}. This compositional sparsity can be further enhanced by introducing prior constraints on features, such as grouping features into concepts or modelling interactions among them. This approach aligns with structured sparsity and hierarchical feature learning \cite{Bach:2010}, which have also been explored in various deep learning studies \cite{ZHANG:2024,Yoon:2017,SCARDAPANE:2017}. \\
Biologically-informed models employ biological knowledge from functional annotation databases to enhance the learning process and improve prediction performance \cite{BINN:2023,PNET:2021}. Many of these approaches are based on neural network architectures, considering pathway annotations as biological information. For example, some of these models employ multi-layer perceptrons (MLPs), where neural connections are modified to incorporate biological pathways. The design of these architectures might be simple, using a single hidden layer \cite{MultiScaleNN:2020,MPVNN:2022} and using a fully connected network associated with the pathway layer \cite{PathDeep:2021}, a sparse coding mechanism with dropout to enhance sparsity effects, along with gene-pathway pruned connections \cite{PASNet:2018,CoxPASNet:2018,MiNet:2019,pathDNN:2020}. Another way of integration is to modify all intermediate layers with pathway information, fitting a sequential neural network structure \cite{PNET:2021,GCSNet:2022,BINN:2023}, or use a parallel fully connected network, incorporating features from all gene features, therefore including also those not associated with pathways \cite{PINNet:2023}. Recently, biologically-informed deep learning models also introduced self-attention mechanisms to the omic-pathway layer \cite{DeepKEGG:2024}, transformers to enhance the interaction between pathways and different data modalities \cite{Pathformer:2024}, or even variational autoencoders that generate a latent data representation, integrating the pathway information into the encoder \cite{AutoSurv:2024}.
All these methods therefore shape the network topology ensuring that functionally-related gene products (or other biological entities) share connections to the same neurons, while pruning connections according to the pathway annotations. 
Another way to exploit pathway information is to transform the input data to reflect pathway relations, enabling the use of neural network architectures designed for non-tabular data. Examples of these architectures are Graph Neural Networks (GNNs) \cite{GCNMAE:2020}, which can represent specific pathways considering the gene-related features as nodes connected according to pathway-specific relationships \cite{PathGNN:2022}, or using pathways as nodes and edges reflecting pathway interactions to be exploited through either graph convolutional layers \cite{PGLCN:2023} or attention mechanisms \cite{GraphPath:2024}. A complementary data transformation strategy involves constructing a two-dimensional “pathway image” that directly encodes gene–pathway associations into a matrix, with gene expression levels represented as "pixel intensities." This format allows standard architectures like CNNs to leverage the structural information provided by pathways for prediction tasks \cite{PathCNN:2021}. Alternatively, these images can be pathway-specific, where gene-related features are ordered according to a similarity metric to position similar features close together \cite{ReGeNNe:2023}. 

All these approaches aim to apply biologically meaningful constraints
able to reduce the model complexity by highlighting relationships that might otherwise remain hidden in raw, high-dimensional datasets. As an example, if a pathway links Gene A and Gene B to a biological function, a pathway-informed neural network ensures that their corresponding input nodes connect to the same subnetwork, preserving their functional context. This enforces sparse connectivity, reducing the number of trainable parameters and improving generalization. This is particularly advantageous when working with high-dimensional, low-sample size data, a common challenge in biomedical research. 

A schematic representation of these concepts is shown in Figure \ref{fig:abstract}. 

Since these algorithms are used also to explain the potential biological impact of specific processes driving an investigated disease, it remains uncertain whether this advantage arises from the biological knowledge itself or as a side effect due to the pathway annotation, which enforces the compositional sparsity exploited by the deep learning algorithms. 
To address this question, we reviewed all the pathway-informed deep learning models we found in the literature and we selected the 20 with publicly available code for systematic evaluation across diverse prediction scenarios. Specifically, we compare their performance against models that use randomized pathway information but preserving the impact of the sparsity on the model. We believe that answering this critical question is essential for guiding future research in the integration of biological priors into machine learning frameworks.
To this aim, in this study we first provide a comprehensive overview of the state-of-the-art of pathway-integrating machine learning models, and then we present a comparison of these techniques using pathway-based vs. randomized sparse information used as prior, evaluating their performance across different prediction tasks.
Finally, we present a workflow to determine whether introducing pathway information could be potentially beneficial for models to be implemented or not.

\section*{Results}

\subsection*{State-of-the-art pathway-informed approaches in deep learning}

\begin{figure}[ht]
 \centering
 \includegraphics[width=\linewidth]{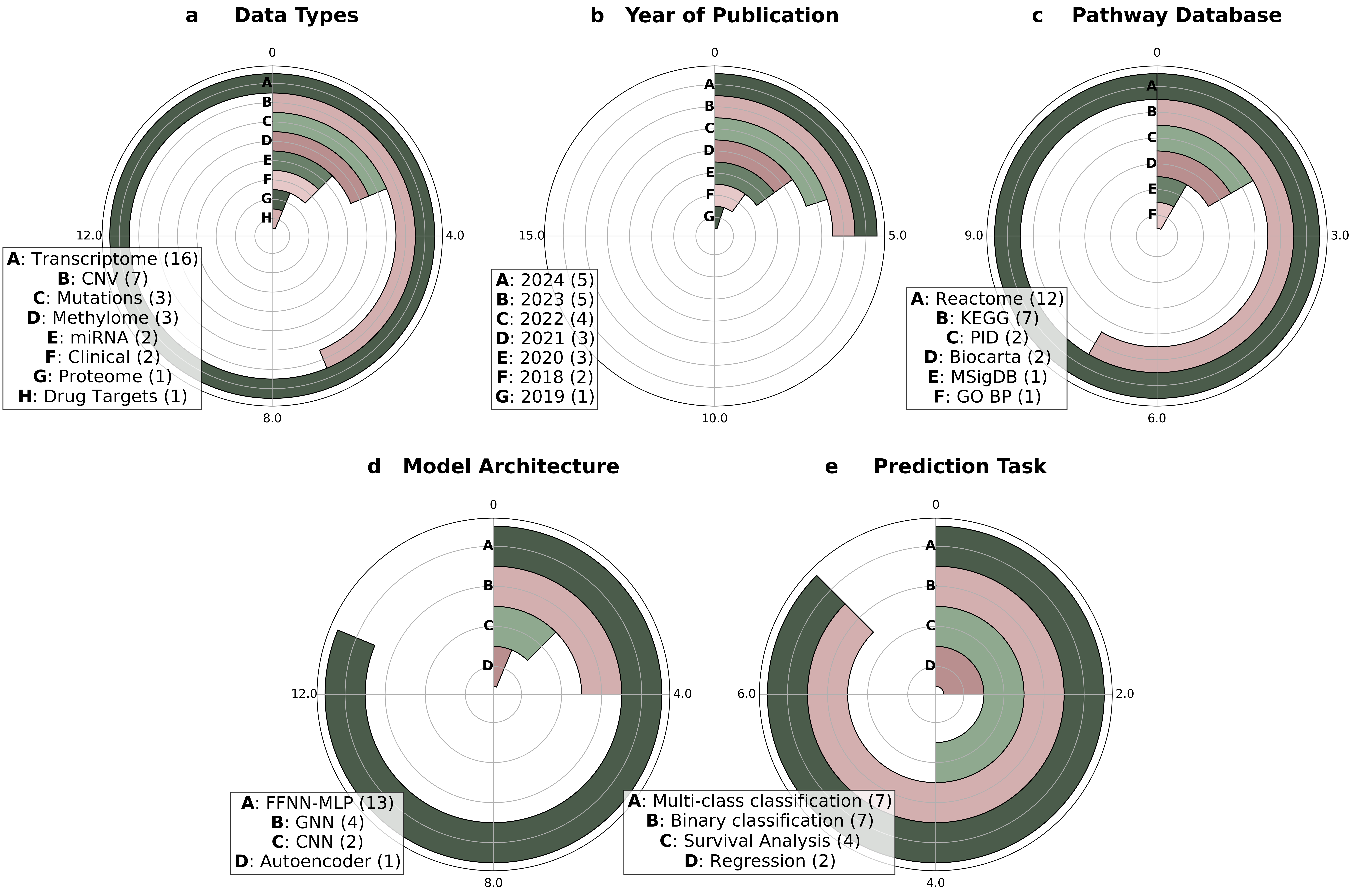}
 \caption{ \textbf{Circular bar plots summarizing characteristics of deep learning models that integrate pathway information}. \\
 The plots show distributions for \textbf{(a)}, Data Types used, \textbf{(b)}, Year of Publication, \textbf{(c)}, Pathway Database sources, \textbf{(d)}, Model Architectures (FFNN-MLP: Feed-Forward Neural Network - Multi-Layer Perceptron, GNN: Graph Neural Network, CNN: Convolutional Neural Network, AE: Autoencoders), and \textbf{(e)}, Prediction Tasks. Each segment’s length corresponds to the count of models within each category.}
 \label{fig:circ_bars}
 \end{figure}

We reviewed all the biologically inspired pathway-based neural network studies we found, examining their methods, assumptions, and results. From these, we selected those with publicly available code to ensure a fair and reproducible comparison using their original implementations.
Table \ref{fig:summary_table} presents a summary of the most recent biologically-informed neural networks able to incorporate the pathway annotations to influence either the model structure or the data organization. As shown also in Figure \ref{fig:circ_bars}, the models address different prediction tasks, ranging from binary and multi-class classification to survival analysis and regression. Table \ref{tab:model_statistics}, Table \ref{tab:mpvnn_statistics} and Table \ref{tab:deepkegg_statistics} in Additional Information provides some statistics of the feature space and the number of samples related to the data used in each study. In terms of pathway annotations used to retrieve the biological information, most of these algorithms rely on Reactome \cite{Reactome} (12 models), followed by KEGG \cite{kegg} (7 models), PID \cite{PID:2009} and Biocarta \cite{Biocarta}(2 models each), and finally GO BP \cite{GO:2000} (used combined with KEGG pathways in 1 model) and MSigDB \cite{MSigDB} (1 model). 
In their respective studies, there are models that exploited the simple associations between gene products and pathways as prior to handling groups of features, while others were able to also include the interactions between gene products associated with the same pathway, and also to include pathway-pathway interactions. 

\subsection*{Pathway-Informed vs. Randomized Models}
\begin{figure}[ht]
 \centering
 \includegraphics[width=\linewidth]{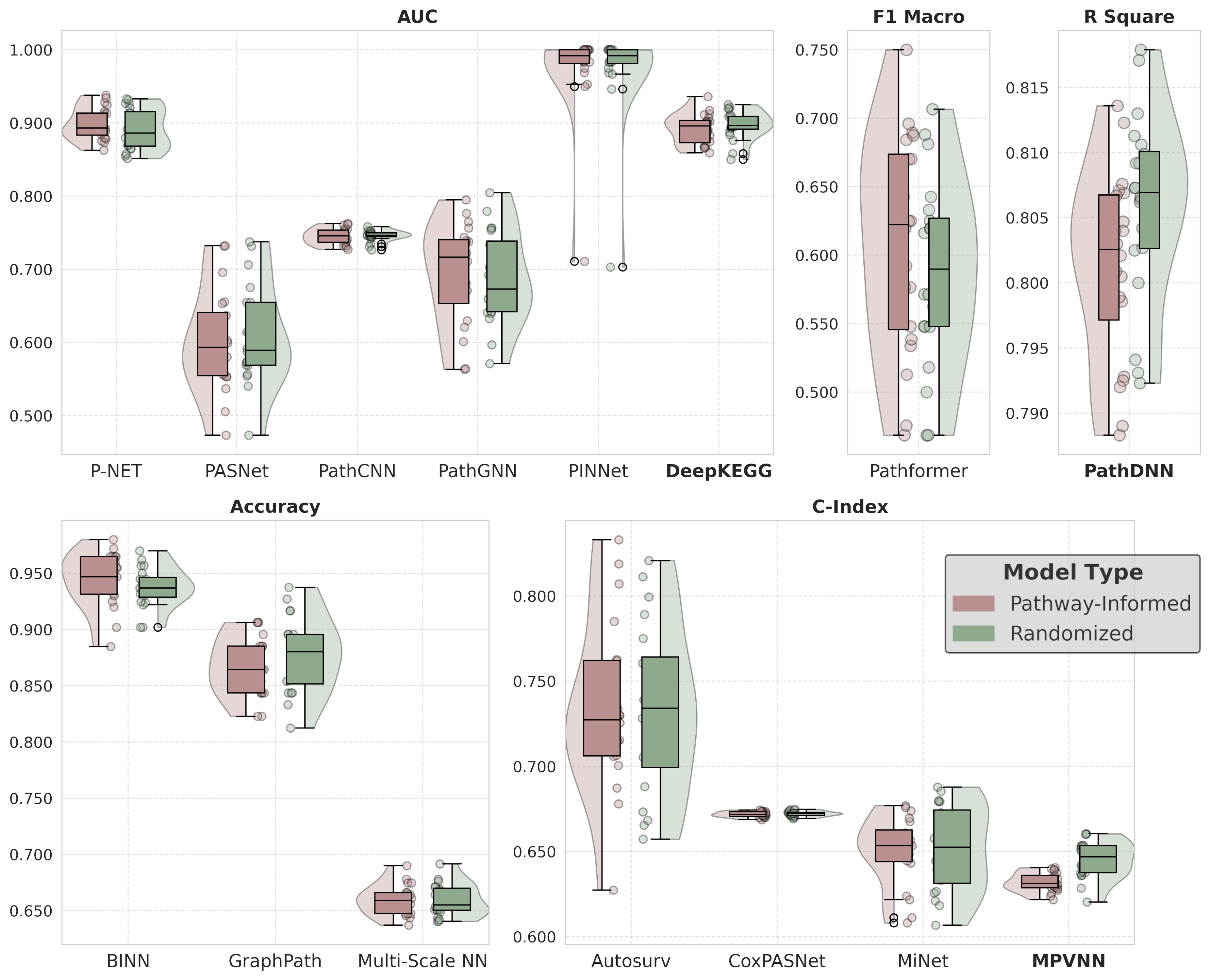}
 \caption{\textbf{Model performance comparison across Accuracy, AUC, C-Index, F1 Macro, and R-Square metrics using violin plots.}\\
 Models are grouped as Pathway-Informed (pink) and Randomized (green). The width reflects the distribution of scores, with central lines for median values and box plots indicating interquartile ranges. Models for which the performance of the randomized version is significantly better than the pathway-informed version are bolded in the x-axis labels.\\
 The results for the MPVNN and DeepKEGG models represent average outcomes across different tumor types considered (detailed findings for each specific tumor type are provided in the Additional Information).}
 \label{fig:results}
 \end{figure}

For each state-of-the-art method reported in Table \ref{fig:summary_table}, we performed a comparison between the original pathway-informed model, that integrates biological priors to guide learning, and its randomized version by replacing these priors with random associations, but preserving network sparsity and structural integrity. For each comparison, 20 independent runs were performed in each model. \\
Figure \ref{fig:results} shows the obtained results, according to the prediction task considered in the original study of each method, therefore stratifying by different performance metrics. Interestingly, for all tested models, the randomized versions performed as well as, or even better than, their biologically-informed counterparts. Notably, for models such as MPVNN, DeepKEGG and PathDNN the randomized versions displayed significantly higher performance. This is supported by the results of statistical tests, including the Kolmogorov-Smirnov and Wilcoxon tests, which consistently indicated that the randomized models outperformed their biologically-informed equivalents. For the remaining models, no significant differences were observed between pathway-informed and randomized versions across all performance metrics, suggesting that incorporating pathway information did not confer a substantial advantage in enhancing model performance for these architectures. \\
To validate that the results were not driven by an unusually favourable random seed, we also generated 30 independent randomizations of the pathway information, with each randomization evaluated across 20 independent runs of the model. Due to computational constraints, this additional trial was conducted on models with feasible runtimes, specifically PINNet, BINN, DeepKEGG, PathCNN and PASNet. Figure \ref{fig:rand_trials} in the Additional Information illustrates that the average performance of the random models using a single seed aligns with the distribution of the average performance obtained from the 30 different randomizations of the pathway data (Kolmogorov-Smirnov test p-value always > 0.05) 
This indicates that the selected randomization seed was not anomalously favourable, but rather shows that the expected variability from randomization is not associated to a “lucky” seed. 

%

Execution times varied across models, with more complex architectures such as PathGNN, Pathformer, and GraphPath requiring significantly longer training durations (even days for single runs). Due to their higher memory demands, models like AutoSurv, GraphPath, and Pathformer were run on a different GPU with greater capacity than the other methods. While this impacted the absolute runtime, execution times remained consistent in order of magnitude. Notably, in some cases, training the more demanding models required several days of continuous computation. Substantial computational effort involved careful hardware optimization and resource allocation to ensure fair comparisons across models. Table \ref{tableres} reports the execution times, approximating the order of magnitude of the extensive resources dedicated to these analyses.





\subsection*{The Role of Sparsity as prior constraint in biologically-informed neural networks}
The previous comparison demonstrated that biologically-inspired neural networks perform equivalently or worse than their randomized counterparts. By construction, our randomized networks preserved the same level of sparsity found in their biologically-informed counterparts. This raises the possibility that the typical sparsity observed in biological systems may be inherently optimal for conveying information and would suggest that biological information influences learning primarily through graph topology rather than explicit pathway annotations.
To investigate this, we compared randomized neural networks at different sparsity levels around those found in biological networks without incorporating biological information. We then evaluated whether sparsity levels derived from biological pathways provided a performance advantage over alternative non-biological sparsity constraints. \\

\begin{figure}[ht]
 \centering
 \includegraphics[width=\linewidth]{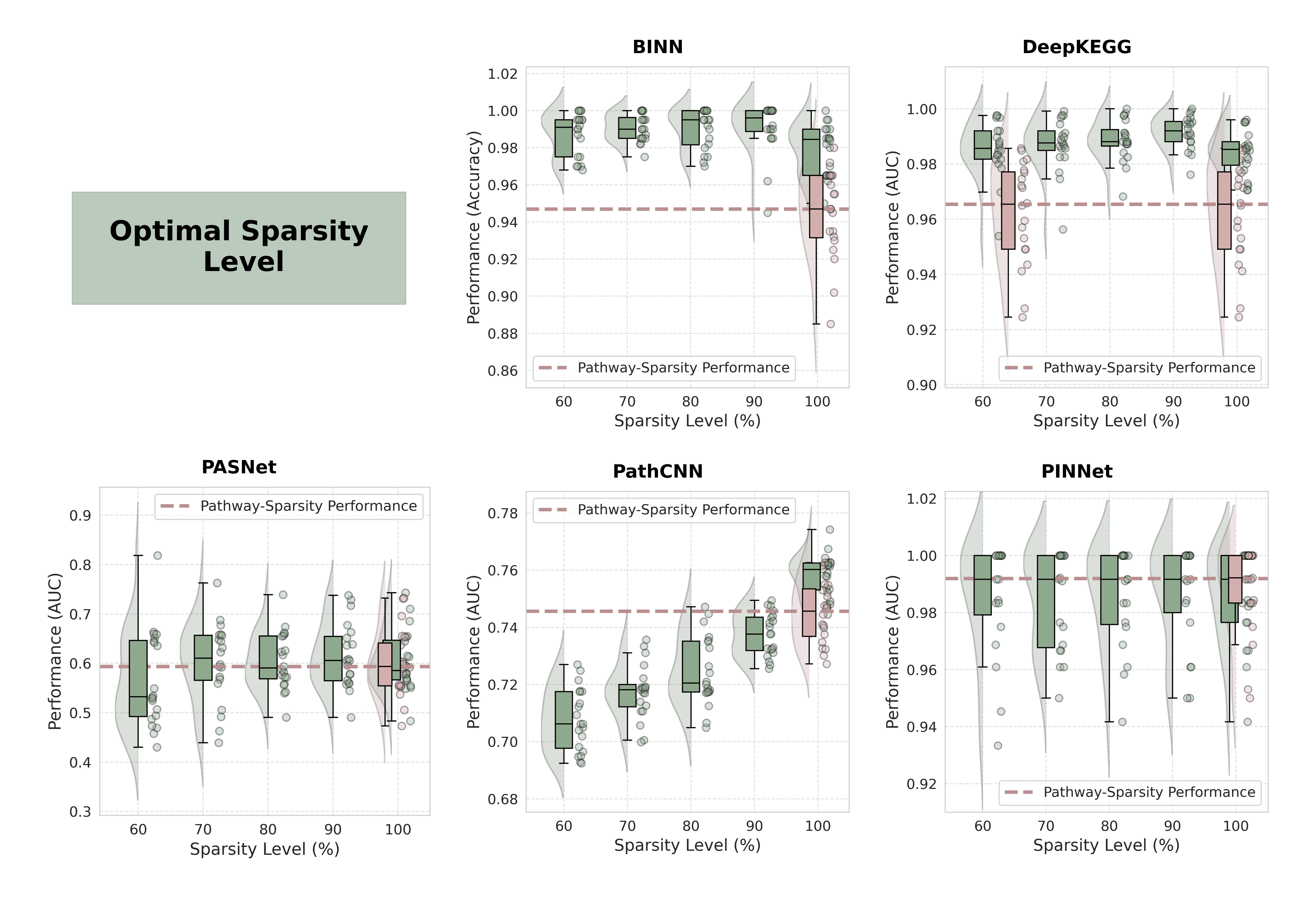}
 \caption{ \textbf{Impact of sparsity on model performance}.
Optimal Sparsity Level: The green boxplots represent the performance (measured as Accuracy or AUC) of each model—BINN, DeepKEGG, PASNet, PathCNN and PINNet—across varying sparsity levels (60\% to 99\%). The pink boxplots indicate performance at the sparsity level induced by pathway information. For DeepKEGG, the pink boxplots are repeated, as the pathway-induced sparsity level varies across omics, ranging from 63.7\% for miRNAs to 98.9\% for mRNAs. In general, boxplots illustrate the distribution of performance across runs, while violin plots provide density estimates. The dashed pink line marks the performance of the pathway-derived sparsity model. Pathway-Induced sparsity levels for all models are reported in Tables \ref{tab:models_sparsity_table}, \ref{tab:deepkegg} and \ref{tab:models_pathway_sparsity_table} in the Additional Information.
}
 \label{fig:main_sparsity}
 \end{figure}

In Figure \ref{fig:main_sparsity}, we report the results for the five neural networks that could be feasibly trained and tested under different conditions: BINN, DeepKEGG, PASNet, PathCNN and PINNet. Comparing different levels of sparsity (ranging from 60\% to 99\%) to the sparsity derived from pathway-based annotations, we found that biologically-induced sparsity led to performance either similar to or significantly lower than the optimal sparsity level. Statistically significant differences favouring non-biologically-induced sparsity were observed only in BINN and DeepKEGG (maximum p-value $1.4 \times 10^{-7}$). Table \ref{tab:models_sparsity_table} in Additional Information presents a table illustrating pathway-induced sparsity levels across models.
These findings suggest that the sparsity characteristic of biological pathways is not necessarily optimal for training neural networks. 

\subsection*{Comparison of Biological Information Extracted by Pathway-Informed Models and Randomized Counterparts}


In order to evaluate whether the interpretability of the underlying biological mechanisms is driven by the integrated pathway information or it can be achieved without its integration, we assessed whether, even in the absence of pathway information, the randomized model could still identify relevant biomarkers by considering the following biologically-informed models: PINNet, DeepKEGG, BINN and PASNet.
A visual representation of the correlation between feature rankings in pathway-informed and randomized models for all four models is provided in the Additional Information at Figure \ref{fig:importance_trials}.
In PINNet, we assessed the relevance of genes identified through SHAP \cite{NIPS2017_7062} by comparing their importance scores to known Alzheimer's disease (AD)-related genes cataloged in the AlzGene \cite{AlzGene:2007} database. Genes listed in AlzGene were classified as AD-related, while all others were considered non-AD-related. The results showed that AD-related genes contributed significantly in both pathway-informed and randomized versions of PINNet, with a p-value < 0.001 in each case, indicating a strong agreement between model predictions and established biological knowledge.
In DeepKEGG, we compared the most important features identified by both model versions to known tumor-related genes from the GeDiPNet \cite{GeDiPNet:2022} database. Among the top 100 ranked features, the pathway-informed model identified 21 tumor-related features, while the randomized model identified 20, demonstrating an almost identical overlap.
For PASNet, where feature names were unavailable, we evaluated the similarity between pathway-informed and randomized models using Spearman’s rank correlation, obtaining a correlation coefficient of 0.4. Given the complexity of the task, this value suggests a moderate to strong alignment between the two versions. The same approach was applied to BINN, where the correlation was 0.56, indicating an even stronger similarity. \\
In general, these findings highlight that models effectively identify disease-relevant biomarkers where applicable, regardless of explicit pathway integration. Moreover, the significant correlation between feature importance rankings in pathway-informed and randomized models further challenges the assumption that biological pathway information is essential for guiding feature selection, as even randomized models recover meaningful biological signals.


\section*{Discussion}

In the present manuscript, we argued, through a comprehensive analysis of multiple learning scenarios, that the performance improvement seen in pathway-informed methodologies might be largely due to the sparsity effect introduced by biological pathway priors rather than the biological relevance of the pathways themselves. Our results showed that randomized models often performed equally well or even outperformed biologically-informed ones across various metrics and datasets, providing strong support for our hypothesis.

This observation is particularly evident in models like MPVNN, DeepKEGG, and PathDNN, in which randomized versions outperformed biologically-informed ones. While biological pathways introduce useful structural sparsity, their actual biological context may not provide additional predictive value. Random sparsification alone proved to be sufficient to obtain the performance gains attributed to biological pathways. These findings suggest that pathway-based models may not always offer a distinct advantage, especially when alternative randomization techniques can induce comparable levels of sparsity.

Further experiments on BINN, DeepKEGG, PASNet, PathCNN and PINNet models reinforce our observations, showing that the choice of a specific seed for pathway randomization is not crucial.

We tested whether removing the most predictive features would reveal a stronger role for pathway information, based on the idea that biological systems may exhibit redundancy and robustness. To assess this, we performed a feature ablation analysis. As shown in Figure~\ref{fig:main_ablation} (Additional Information), performance remained comparable between the pathway-informed and randomized models at every stage of feature removal. This indicates that pathway information was not simply masked by highly predictive biomarkers. Across all models and ablation steps, the differences in performance were not statistically significant, further supporting the conclusion that pathway priors offer limited benefit over randomization in terms of predictive accuracy.

Furthermore, our analysis suggests that the optimal level of sparsity does not necessarily overlap with the sparsity imposed by biological pathways. For instance, models such as BINN and DeepKEGG exhibited significant differences between pathway-induced sparsity and the level that yielded the best predictive performance. This highlights the importance of treating sparsity as a tunable hyperparameter rather than a fixed property dictated by biological priors.

Regarding model interpretability, both the pathway-informed and randomized networks yield similar outcomes, with each being capable of identifying relevant biomarkers for the disease under investigation where applicable. Moreover, the feature importance rankings derived from pathway-informed and randomized models exhibit a significant degree of correlation, suggesting that even in the absence of explicit biological priors, randomized models can still capture key features associated with the problem under consideration.



\begin{figure}[ht]
 \centering
 \includegraphics[width=\linewidth]{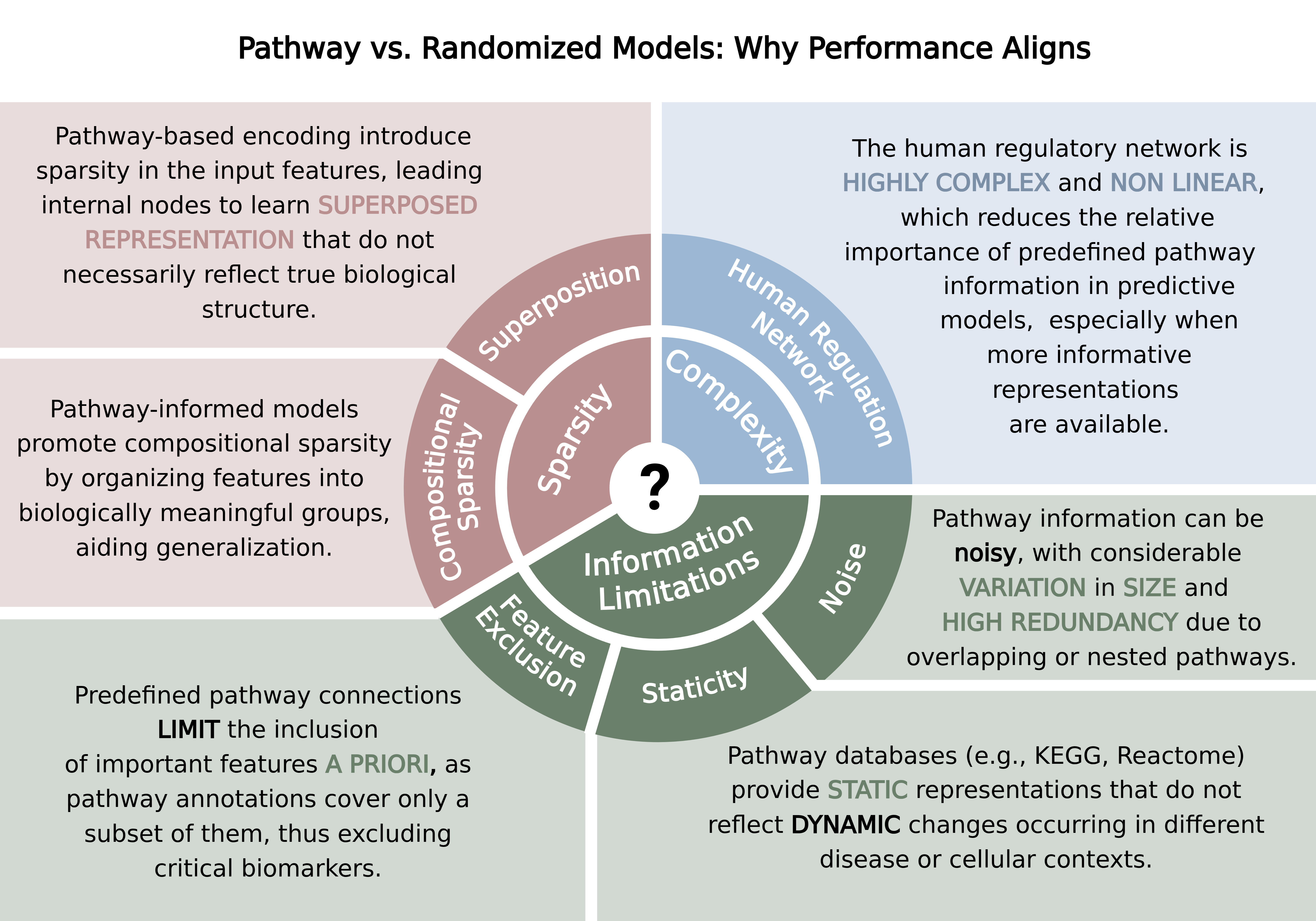}
    \caption{
    \textbf{Hypothetical causes for the alignment in performance between pathway-informed and randomized models.}
    Despite integrating biological knowledge, randomized models often perform comparably or better with respect to models incorporating pathway information. This figure summarizes several hypothetical factors that may contribute to explain this phenomenon.}
 \label{fig:perf_causes}
 \end{figure}

Several factors may explain why pathway integration does not improve performance beyond the sparsity effect it produces. A schematic illustration of the proposed motivations is presented in Figure \ref{fig:perf_causes}. One possibility is that predefined pathway connections limit the inclusion of important genes, as pathway annotations cover only a subset of genes, potentially excluding critical biomarkers. Furthermore, it is possible that the imposed pathway-based encoding leads to sparsity in the input features, causing internal nodes to develop superposed representations (representations that combine multiple unrelated signals) that do not necessarily align with the underlying biological structure~\cite{elhage2022toy}. This could explain why pathway-informed models do not outperform their randomized counterparts and why their explanations may be less useful for interpretation. This issue is particularly relevant for the examined approaches, where important genes may be overlooked simply because they are not annotated in the pathway databases. Future studies could explore the integration of protein-protein interaction (PPI) networks, which encompass a broader range of genes, to assess whether these networks can improve the benefits of sparsity alone.

Moreover, pathway information from sources like Reactome or KEGG is static, failing to reflect the dynamic nature of biological processes that often evolve in the context of diseases. Pathways are not fixed entities; they can change depending on cellular states or environmental conditions. Relying on static representations may, therefore, oversimplify the complexity of disease mechanisms, limiting the effectiveness of pathway-informed models.

Additionally, the human regulatory network is highly complex and nonlinear, which may render pathway information less crucial for predictive models compared to other, more informative data types. The static and incomplete nature of current pathway annotations could be overshadowed by other forms of biological information that capture more dynamic aspects of the system.


In this regard, it would be valuable to explore whether pathway information plays a more predictive role in simpler organisms, such as bacteria, where regulatory networks are less complex. In such cases, pathways might provide benefits beyond the sparsity effect, leading to more accurate models and helping clarify whether the limited utility of pathway information in human models is due to system complexity or the limitations of current pathway datasets. This remains beyond the scope of the present study.

In conclusion, our findings open up important questions about the role of biological priors in deep learning models. While sparsity remains a key factor in improving model performance, our study suggests that sparsity alone, without the inclusion of biological knowledge, can often be sufficient. This could lead to a shift in how biologically-informed models are developed, focusing more on structural advantages like sparsity rather than on the incorporation of specific biological data.


Future work should consistently validate pathway integration by comparing model performance with its randomized counterpart. Such comparisons will ensure that the integration of biological information offers benefits beyond sparsity and genuinely enhances predictive capabilities.

\section*{Methods}

\begin{figure}[ht]
 \centering
 \includegraphics[width=\linewidth]{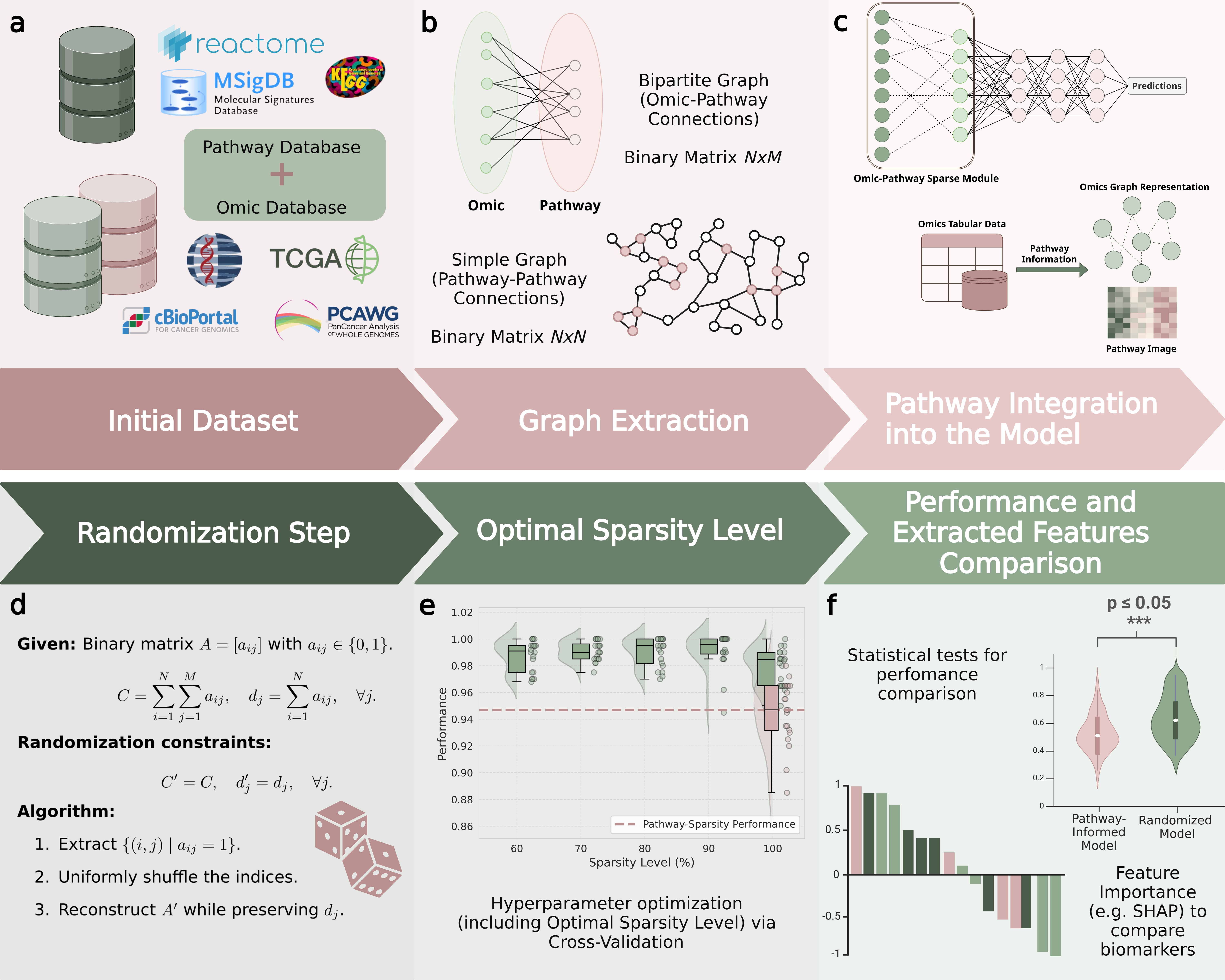}
    \caption{
    \textbf{Guidelines for Integrating Biological Pathways into Predictive Models with Proper Benchmarking.}
    This figure outlines a principled workflow for incorporating biological pathway knowledge into omics-based predictive models while ensuring robust validation against randomized baselines. 
    \textbf{(a)} Datasets from pathway (e.g., Reactome, KEGG) and omics sources (e.g., TCGA, PCAWG) are combined to build a bipartite graph linking omic features to pathways or a simple graph linking pathways to each another. 
    \textbf{(b)} The graph are encoded as a binary matrix either representing feature-to-pathway or pathway-to-pathway associations. 
    \textbf{(c)} The graph structure is embedded into the model via a sparse omic-pathway module that enriches standard omics data with biologically-informed connectivity or by modifying the structure of input data (e.g. in GNNs and CNNs based models). 
    \textbf{(d)} To assess the added value of true biological structure, a randomization step permutes pathway connections while preserving degree distributions, ensuring fair comparison. 
    \textbf{(e)} Optional: 
    Optimize the sparsity of the omic-pathway graph to achieve better predictive performance. This is done using a cross-validation framework. In this step, the original degree distribution constraint is relaxed, allowing for a more flexible exploration of graph structures that may enhance model's accuracy.
    \textbf{(f)} Statistical analyses and feature attribution methods (e.g., SHAP) are employed to compare model performance and feature relevance between biologically-informed and randomized counterparts.
    This whole approach enables rigorous validation of pathway integration, ensuring that observed improvements are due to meaningful biological priors.
    }
 \label{fig:supp-guidelines}
 \end{figure}

Figure \ref{fig:supp-guidelines} outlines a practical set of guidelines for integrating biological pathway information into omics-based predictive models. Serving as a step-by-step workflow, it demonstrates how to combine pathway and omics data, encode these associations into graph representations, embed the resulting structures into neural network architectures, and benchmark performance rigorously against randomized baselines. These guidelines offer a clear framework that summarizes the following Methods section, helping researchers identify whether performance gains come from biological priors or from beneficial sparsity effects.

\subsection*{Randomization procedure}
\begin{figure}[ht]
 \centering
 \includegraphics[width=\linewidth]{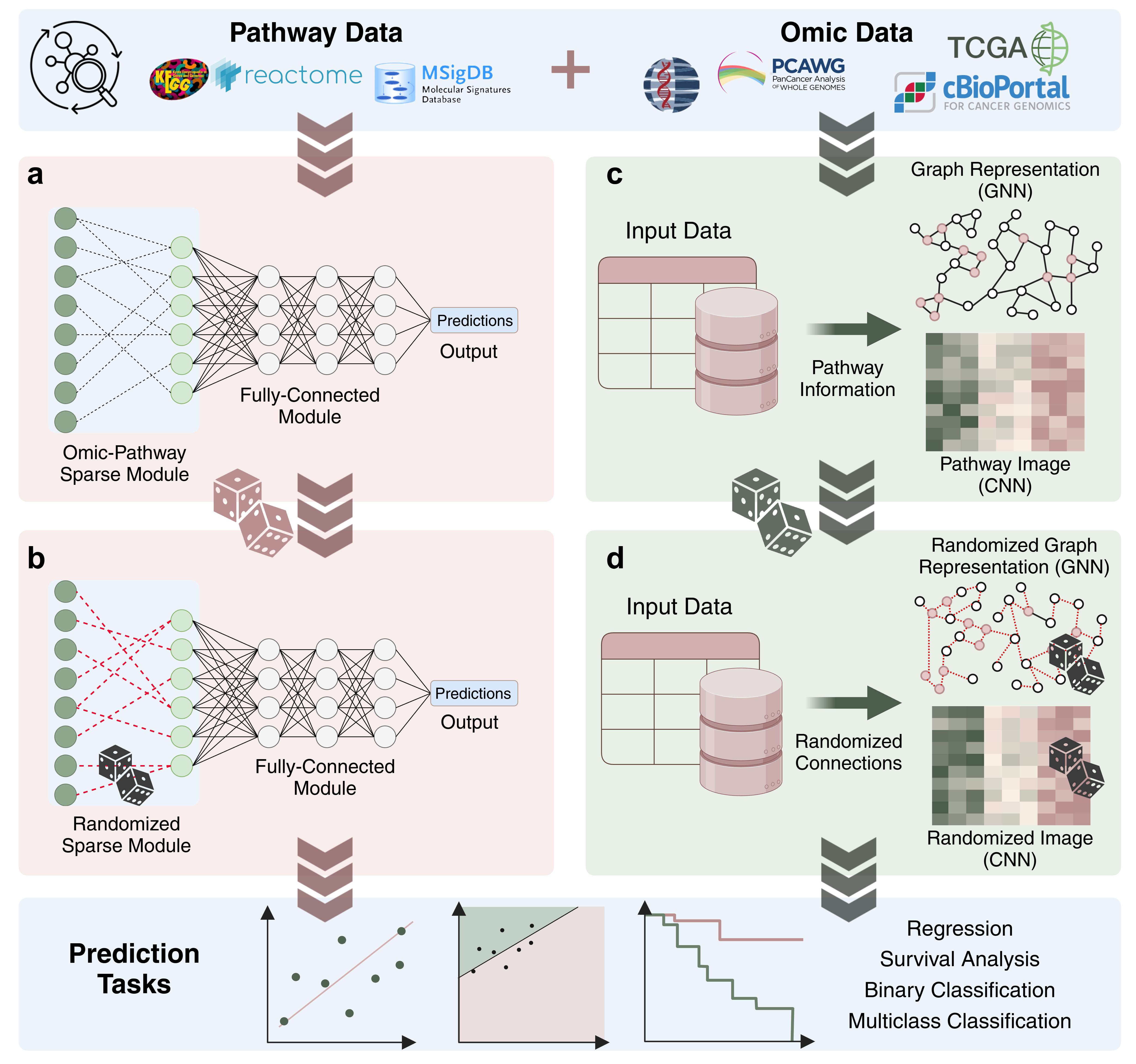}
 \caption{\textbf{Schematic representation of pathway integration approaches in neural networks for omics data and their relative randomization.}\\
 Pathway information can be incorporated in two ways (Panels a and c):  \textbf{(a)} A neural network utilizing pathway information by enforcing structured connections, introducing sparsity in the model. \textbf{(b)} A randomized counterpart where connections are introduced without explicit pathway constraints allows for an alternative exploration of the data structure. \textbf{(c)} A data transformation strategy that incorporates pathway information to convert tabular omics data into graphs or images. \textbf{(d)} A randomized data transformation approach that generates graphs or images through a randomization procedure rather than predefined pathway structures. }
 \label{fig:abstract}
 \end{figure}

The process of randomizing pathway information entails generating a null model by permuting pathway-based associations. In the approach illustrated in Figure \ref{fig:abstract}, panel (a), neuron connections within neural networks are replaced with random ones, while maintaining the same number of connections per neuron. This preserves the sparsity effect inherent to pathway integration within the models. Similarly, in the modality shown in Figure \ref{fig:abstract}, panel (c), randomization involves transforming tabular data into structured data by substituting the original pathway priors. Specifically, in Graph Neural Networks, this is achieved by introducing random connections among the nodes in the input graphs. For Convolutional Neural Networks, the randomization step consists of constructing a “pathway image” by assigning random omics entities to each pathway. Again, in both cases, the number of connections in the network or the number of omics entities per pathway is preserved to maintain the same level of sparsity that was achieved through the use of biological priors. This ensures that the randomization process mirrors the structural characteristics of the original models, preserving the sparsity effects while eliminating the biological relevance of the pathway information. \\

\subsection*{Hyperparameter selection}

After randomizing the model structures and input data, both the biologically-informed models and their randomized counterparts were run to compare performance. The experimental procedure involved generating 20 distinct training and testing sets for each model, using an 80/20 split. Each split was created by randomly dividing the samples, with stratification applied according to the task-specific labels when necessary. When optimal hyperparameter values were specified in the models' repositories, we used the same values for the randomized models. Otherwise, we optimized the hyperparameters for the biologically-informed models using a cross-validation procedure on the training set and consequently evaluated on the test set.

\subsection*{Extended Analysis of Pathway-Informed Models}
In addition to comparing the performance of randomized and pathway-informed model variants, further analyses were performed on PINNet, the fastest model to run, along with four other models: BINN, DeepKEGG, PASNet and PathCNN. \\

\subsubsection*{Randomization trials}

In this analysis, 30 different randomizations of the pathway information were generated, and for each randomization, the model was run 20 times. This was done to ensure that the results obtained with a single randomization were not due to a particularly favourable random seed.\\

 \subsubsection*{Optimal Sparsity Level}

We questioned whether the biological pathways contributed information primarily through the optimal level of sparsity, rather than through the specific connections or features they introduced. In other words, the biological signal provided by pathways might lie not in the precise connections retained, but in the overall number of connections within the neural network. To investigate this aspect, we first constructed a sparse neural network where sparsity was defined based on the number of pathway connections and compared its performance to a fully connected model. Subsequently, we extended the analysis by examining models with varying levels of sparsity, ranging from 60\% to 99\% pruned connections, along with a model where the sparsity level was dictated by biological pathway-derived connections. The sparsity thresholds were chosen by varying around the biological sparsity induced by pathway priors. As shown in Table \ref{tab:models_sparsity_table} and \ref{tab:deepkegg}, the most common level of pathway-induced sparsity was approximately 97–99\%, while for miRNA in the DeepKEGG model, it was around 59–66\%. Statistical comparisons were conducted to determine whether pathway-informed sparsity provided an advantage over arbitrary levels of sparsity. The purpose of this procedure is to compare, in a post hoc analysis on the test set, the results obtained from trials conducted at different sparsity levels to assess whether they yield comparable outcomes. However, if the goal was to determine the optimal sparsity level for a given model, it should be treated as a standard hyperparameter and hence identified during the validation phase.\\

\subsubsection*{Comparison of Biological Information Extracted by Pathway-Informed Models and Randomized Counterparts}

Finally, we investigated models interpretability to assess whether, even in the absence of pathway information, the randomized model could still identify relevant biomarkers for the disease under study. In PINNet, this was done by comparing the importance scores of the genes identified through SHAP with known AD-related genes. Specifically, the genes cataloged in the AlzGene database were considered to be AD-related, the remaining as not AD-related. In DeepKEGG, the most important features identified by both pathway-informed and randomized models were compared to known tumor-related genes from the GeDiPNet database. For PASNet, where feature names were unavailable, we compared the ranked importance of features in pathway-informed and randomized models using Spearman’s rank correlation. The same correlation-based approach was applied to BINN.\\

\subsubsection*{Features ablation study}

A gradual feature ablation study was conducted on the models to assess whether removing key features — identified as important for prediction - would highlight the role of pathways. To identify and discard highly discriminative features from each dataset, we employed a Mann-Whitney U test-based approach. The goal was to determine whether the influence of pathways was being overshadowed by the contribution of highly predictive features. After each set of features was removed, the performance of the pathway-informed and randomized model versions was compared again. \\

\subsection*{Comparison between biological-informed and random counterparts}
Comparisons among result distributions were conducted using two statistical tests: the paired samples Wilcoxon test and the Kolmogorov-Smirnov test. Both two-sided and one-sided alternatives were evaluated, with statistical significance set at a threshold of $p < 0.05$. \\

Unfortunately, it was not possible to perform the performance comparison for models GCN-MAE and GCS-Net due to the unavailability of the code in their respective GitHub repositories. Additionally, models PathDeep, ReGeNNe, and PGLCN could not be included in the analysis because the necessary data for making predictions were not available. \\
The analyses were executed on an NVIDIA GeForce RTX 4070 Max-Q GPU with 8 GB of memory. For models with higher memory demands (e.g., AutoSurv, GraphPath, and Pathformer), a Tesla V100 SXM2 GPU with 32 GB of memory was utilized.

\section*{Data \& Code Availability}

The code used for the pathway connections randomization procedure can be found at \\
\href{https://github.com/compbiomed-unito/Pathway_Randomization}{https://github.com/compbiomed-unito/Pathway\_Randomization}. \\This repository provides tools for pathway randomization in neural networks for omics data analysis, including functions to shuffle pathway connections while preserving specific constraints (e.g. desired sparsity levels). \\
Code and datasets used to train the specific models were obtained from their respective repositories. A list of the models along with the links to their repositories can be found in the Additional Information. \\

\bibliographystyle{naturemag-doi}  
\bibliography{sample}

\begin{thebibliography}{10}

\bibitem{GO:2000}
Michael Ashburner, Catherine~A. Ball, Judith~A. Blake, David Botstein, Heather
  Butler, J.~Michael Cherry, Allan~P. Davis, Kara Dolinski, Selina~S. Dwight,
  Janan~T. Eppig, Midori~A. Harris, David~P. Hill, Lori Issel-Tarver, Andrew
  Kasarskis, Suzanna Lewis, John~C. Matese, Joel~E. Richardson, Martin
  Ringwald, Gerald~M. Rubin, and Gavin Sherlock.
\newblock Gene ontology: tool for the unification of biology.
\newblock {\em Nature Genetics}, 25:25--29, 2000.

\bibitem{Bach:2010}
Francis Bach.
\newblock Structured sparsity-inducing norms through submodular functions.
\newblock {\em arXiv preprint}, arXiv:1008.4220, 2010.

\bibitem{AlzGene:2007}
L.~Bertram, M.~McQueen, K.~Mullin, D.~Blacker, and R.~E. Tanzi.
\newblock Systematic meta-analyses of alzheimer disease genetic association
  studies: the alzgene database.
\newblock {\em Nature Genetics}, 39:17--23, 2007.

\bibitem{pathDNN:2020}
L.~Deng, Y.~Cai, W.~Zhang, W.~Yang, B.~Gao, and H.~Liu.
\newblock Pathway-guided deep neural network toward interpretable and
  predictive modeling of drug sensitivity.
\newblock {\em Journal of Chemical Information and Modeling}, 2020.

\bibitem{elhage2022toy}
Nelson Elhage, Tristan Hume, Catherine Olsson, Nicholas Schiefer, Tom Henighan,
  Shauna Kravec, Zac Hatfield-Dodds, Robert Lasenby, Dawn Drain, Carol Chen,
  Roger Grosse, Sam McCandlish, Jared Kaplan, Dario Amodei, Martin Wattenberg,
  and Christopher Olah.
\newblock Toy models of superposition.
\newblock {\em arXiv preprint arXiv:2209.10652}, 2022.

\bibitem{PNET:2021}
H.A. Elmarakeby, J.~Hwang, R.~Arafeh, J.~Crowdis, D.and AlDubayan S. H.and
  Salari~K. Gang, S.and~Liu, S.~Kregel, C.~Richter, T.~E. Arnoff, J.~Park,
  W.~C. Hahn, and E.~M. Van~Allen.
\newblock Biologically informed deep neural network for prostate cancer
  discovery.
\newblock {\em Nature}, 2021.

\bibitem{MultiScaleNN:2020}
T.~Gaudelet, N.~Malod-Dognin, J.~Sànchez-Valle, V.~Pancaldi, A.~Valencia, and
  N.~Pržulj.
\newblock Unveiling new disease, pathway, and gene associations via multi-scale
  neural network.
\newblock {\em PLOS One}, 2020.

\bibitem{PASNet:2018}
J.~Hao, Y.~Kim, T.K. Kim, and M.~Kang.
\newblock Pasnet: pathway-associated sparse deep neural network for prognosis
  prediction from high-throughput data.
\newblock {\em BMC Bioinformatics}, 2018.

\bibitem{CoxPASNet:2018}
J.~Hao, Y.~Kim, T.~Mallavarapu, J.~H. Oh, and M.~Kang.
\newblock Cox-pasnet: An artificial neural network for predicting prognosis in
  cancer patients based on pathway-associated sparse deep neural networks.
\newblock In {\em IEEE International Conference on Bioinformatics and
  Biomedicine (BIBM)}, pages 381--386, 2018.

\bibitem{MiNet:2019}
J.~Hao, M.~Masum, J.~H. Oh, and M.~Kang.
\newblock Gene- and pathway-based deep neural network for multi-omics data
  integration to predict cancer survival outcomes.
\newblock In {\em Bioinformatics Research and Applications}, pages 113--124,
  Cham, 2019. Springer International Publishing.

\bibitem{BINN:2023}
E.~Hartman, A.M. Scott, C.~Karlsson, and et~al.
\newblock Interpreting biologically informed neural networks for enhanced
  proteomic biomarker discovery and pathway analysis.
\newblock {\em Nature Communications}, 14, 2023.

\bibitem{hastie2015statistical}
Trevor Hastie, Robert Tibshirani, and Martin Wainwright.
\newblock {\em Statistical Learning with Sparsity: The Lasso and
  Generalizations}.
\newblock Chapman \& Hall/CRC, 2015.

\bibitem{Hoefler:2021}
T.~Hoefler, D.~Alistarh, T.~Ben-Nun, N.~Dryden, and A.~Peste.
\newblock Sparsity in deep learning: pruning and growth for efficient inference
  and training in neural networks.
\newblock {\em Journal of Machine Learning Research}, 2021.

\bibitem{GCSNet:2022}
J.~Hu, W.~Yu, Y.~Dai, C.~Liu, Y.~Wang, and Q.~Wu.
\newblock A deep neural network for gastric cancer prognosis prediction based
  on biological information pathways.
\newblock {\em Journal of Oncology}, 2022.

\bibitem{AutoSurv:2024}
L.~Jiang, C.~Xu, Y.~Bai, A.~Liu, Y.~Gong, Y.~Wang, and H.~Deng.
\newblock Autosurv: interpretable deep learning framework for cancer survival
  analysis incorporating clinical and multi-omics data.
\newblock {\em npj Precision Oncology}, 2024.

\bibitem{kegg}
Minoru Kanehisa, Miho Furumichi, Yoko Sato, Masayuki Kawashima, and Mari
  Ishiguro-Watanabe.
\newblock Kegg for taxonomy-based analysis of pathways and genomes.
\newblock {\em Nucleic Acids Research}, 51(D1):D587--D592, 10 2023.

\bibitem{PINNet:2023}
H.~Kim and H.~Lee.
\newblock Pinnet: a deep neural network with pathway prior knowledge for
  alzheimer's disease.
\newblock {\em Frontiers in Aging Neuroscience}, 2023.

\bibitem{GeDiPNet:2022}
I.~Kundu, M.~Sharma, R.~S. Barai, K.~Pokar, and S.~Idicula-Thomas.
\newblock Gedipnet: Online resource of curated gene-disease associations for
  polypharmacological targets discovery.
\newblock {\em Genes \& Diseases}, 2023.

\bibitem{DeepKEGG:2024}
W.~Lan, H.~Liao, Q.~Chen, L.~Zhu, Y.~Pan, and Y.~P. Chen.
\newblock Deepkegg: a multi-omics data integration framework with biological
  insights for cancer recurrence prediction and biomarker discovery.
\newblock {\em Briefings in Bioinformatics}, 2024.

\bibitem{GCNMAE:2020}
S.~Lee, S.~Lim, T.~Lee, I.~Sung, and S.~Kim.
\newblock Cancer subtype classification and modeling by pathway attention and
  propagation.
\newblock {\em Bioinformatics}, 2020.

\bibitem{PathGNN:2022}
B.~Liang, H.~Gong, L.~Lu, and J.~Xu.
\newblock Risk stratification and pathway analysis based on graph neural
  network and interpretable algorithm.
\newblock {\em BMC Bioinformatics}, 2022.

\bibitem{MSigDB}
Arthur Liberzon, Aravind Subramanian, Reid Pinchback, Helga Thorvaldsdóttir,
  Pablo Tamayo, and Jill~P. Mesirov.
\newblock Molecular signatures database (msigdb) 3.0.
\newblock {\em Bioinformatics}, 27(12):1739--1740, 2011.

\bibitem{PGLCN:2023}
C.~Liu, A.~H. Wan, H.~Liang, L.~Sun, J.~Li, R.~Yang, Q.~Li, R.~Wu, K.~Hu,
  Y.~Yang, S.~Cai, G.~Wan, and W.~He.
\newblock Biological informed graph neural network for tumor mutation burden
  prediction and immunotherapy-related pathway analysis in gastric cancer.
\newblock {\em Computational and Structural Biotechnology Journal}, 2023.

\bibitem{Pathformer:2024}
X.~Liu, Y.~Tao, Z.~Cai, P.~Bao, H.~Ma, K.~Li, M.~Li, Y.~Zhu, and Z.~H. Lu.
\newblock Pathformer: a biological pathway informed transformer for disease
  diagnosis and prognosis using multi-omics data.
\newblock {\em Bioinformatics}, 2024.

\bibitem{NIPS2017_7062}
Scott~M Lundberg and Su-In Lee.
\newblock A unified approach to interpreting model predictions.
\newblock In I.~Guyon, U.~V. Luxburg, S.~Bengio, H.~Wallach, R.~Fergus,
  S.~Vishwanathan, and R.~Garnett, editors, {\em Advances in Neural Information
  Processing Systems 30}, pages 4765--4774. Curran Associates, Inc., 2017.

\bibitem{GraphPath:2024}
T.~Ma and J.~Wang.
\newblock Graphpath: a graph attention model for molecular stratification with
  interpretability based on the pathway–pathway interaction network.
\newblock {\em Bioinformatics}, 2024.

\bibitem{Reactome}
Marija Milacic, Deidre Beavers, Patrick Conley, Chuqiao Gong, Marc Gillespie,
  Johannes Griss, Robin Haw, Bijay Jassal, Lisa Matthews, Bruce May, Robert
  Petryszak, Eliot Ragueneau, Karen Rothfels, Cristoffer Sevilla, Veronica
  Shamovsky, Ralf Stephan, Krishna Tiwari, Thawfeek Varusai, Joel Weiser, Adam
  Wright, Guanming Wu, Lincoln Stein, Henning Hermjakob, and Peter
  D’Eustachio.
\newblock The reactome pathway knowledgebase 2024.
\newblock {\em Nucleic Acids Research}, 52(D1):D672--D678, 2023.

\bibitem{Biocarta}
Darryl Nishimura.
\newblock Biocarta.
\newblock {\em Biotech Software \& Internet Report}, 2004.

\bibitem{PathCNN:2021}
J.~H. Oh, W.~Choi, E.~Ko, M.~Kang, A.~Tannenbaum, and J.~O. Deasy.
\newblock Pathcnn: interpretable convolutional neural networks for survival
  prediction and pathway analysis applied to glioblastoma.
\newblock {\em Bioinformatics}, 2021.

\bibitem{PathDeep:2021}
S.~Park, E.~Huang, and T.~Ahn.
\newblock Classification and functional analysis between cancer and normal
  tissues using explainable pathway deep learning through rna-sequencing gene
  expression.
\newblock {\em International Journal of Molecular Sciences}, 2021.

\bibitem{Poggio:2022}
T.~Poggio.
\newblock How deep sparse networks avoid the curse of dimensionality:
  Efficiently computable functions are compositionally sparse.
\newblock Technical Report CBMM Memo 118, Center for Brains, Minds and Machines
  (CBMM), 2022.
\newblock \url{https://hdl.handle.net/1721.1/145776}.

\bibitem{poggio2024compositional}
Tomaso Poggio and Maia Fraser.
\newblock Compositional sparsity of learnable functions.
\newblock {\em Bulletin of the American Mathematical Society}, 2024.

\bibitem{MPVNN:2022}
G.~G. Roy, N.~Geard, K.~Verspoor, and S.~He.
\newblock Mpvnn: Mutated pathway visible neural network architecture for
  interpretable prediction of cancer-specific survival risk.
\newblock {\em Bioinformatics}, 2022.

\bibitem{SCARDAPANE:2017}
Simone Scardapane, Danilo Comminiello, Amir Hussain, and Aurelio Uncini.
\newblock Group sparse regularization for deep neural networks.
\newblock {\em Neurocomputing}, 241:81--89, 2017.

\bibitem{PID:2009}
Carl~F. Schaefer, Karthik Anthony, Stephen Krupa, Jonathan Buchoff, Michael
  Day, Tom Hannay, and Kenneth~H. Buetow.
\newblock Pid: the pathway interaction database.
\newblock {\em Nucleic Acids Research}, 37:D674--D679, 2009.

\bibitem{ReGeNNe:2023}
D.~Sharma and W.~Xu.
\newblock Regenne: genetic pathway-based deep neural network using canonical
  correlation regularizer for disease prediction.
\newblock {\em Bioinformatics}, 2023.

\bibitem{wen2016learning}
Wei Wen, Chunpeng Wu, Yandan Wang, Yiran Chen, and Hai Li.
\newblock Learning structured sparsity in deep neural networks.
\newblock In {\em Proceedings of the 30th International Conference on Neural
  Information Processing Systems}, volume~30, pages 2082--2090, 2016.

\bibitem{Yoon:2017}
Jaehong Yoon and Sung~Ju Hwang.
\newblock Combined group and exclusive sparsity for deep neural networks.
\newblock In Doina Precup and Yee~Whye Teh, editors, {\em Proceedings of the
  34th International Conference on Machine Learning}, volume~70 of {\em
  Proceedings of Machine Learning Research}, pages 3958--3966. PMLR, 06--11 Aug
  2017.

\bibitem{ZHANG:2024}
Xin Zhang and Junlong Zhao.
\newblock Group variable selection via group sparse neural network.
\newblock {\em Computational Statistics \& Data Analysis}, 192:107911, 2024.

\end{thebibliography}

\section*{Author Information}
\vspace{1em}
\subsection*{Authors and Affiliations}
\vspace{1em}
\begin{flushleft}
    \textbf{Computational Biomedicine Unit, Department of Medical Sciences, University of Torino, Torino, IT}\\[0.5em]
    Isabella Caranzano \\
    Corrado Pancotti \\
    Cesare Rollo \\
    Flavio Sartori \\
    Piero Fariselli \\
    Tiziana Sanavia \\
\end{flushleft}

\begin{flushleft}
    \textbf{Department of Computer Science and Technology, University of Cambridge, Cambridge, UK}\\[0.5em]
    Pietro Liò \\
\end{flushleft}

\subsection*{Contributions}
\vspace{1em}
I.C., P.F. and T.S. conceived the experiments, I.C. conducted the experiments and analysed the results. I.C., P.F. P.L., T.S., C.P. and C.R. collectively discussed the results and contributed to the final manuscript.

\subsection*{Corresponding author}
\vspace{1em}
Piero Fariselli

\section*{Competing interests} 


The authors declare no competing interests.

\begin{longtable}{|>{\columncolor[HTML]{8FA98F}}p{1.9cm}|p{0.7cm}|p{2.5cm}|p{2.5cm}|p{1.5cm}|p{0.8cm}|p{1cm}|p{1.4cm}|}
    
\hline
\rowcolor[HTML]{D3AFAF}
\textbf{Model} & \textbf{Year} & \textbf{Journal} & \textbf{Prediction Task}  & \textbf{Pathway Source} & \textbf{Code} & \textbf{Data} & \textbf{Data Type} \\ 
\hline
\endfirsthead

\hline
\multicolumn{8}{c}{{\bfseries Table \ref{fig:summary_table} (Continued)}} \\ 
\hline
\rowcolor[HTML]{D3AFAF}
\textbf{Model} & \textbf{Year} & \textbf{Journal} & \textbf{Prediction Task}& \textbf{Pathway Source} & \textbf{Code} & \textbf{Data} & \textbf{Data Type} \\ 
\hline
\endhead

\hline
\multicolumn{8}{r}{{Continued on next page}} \\ 
\endfoot

\hline
\caption{\textbf{Overview of deep learning models integrating pathway information for various prediction tasks.}\\
Models are categorized by year of publication, journal, prediction task, pathway source, code availability, data availability, and data type.} 
\label{fig:summary_table} \\
\endlastfoot
\hline
PASNet & 2018 & BMC Bioinformatics & Binary classification Long-term VS Short-term survival & Reactome & \href{https://github.com/DataX-JieHao/PASNet}{\checkmark} & \checkmark & Gene expression \\ \hline
Cox-PASNet & 2018 & IEEE International Conference on Bioinformatics and Biomedicine (BIBM) 2018 & Survival Analysis & Reactome & \href{https://github.com/DataX-JieHao/Cox-PASNet}{\checkmark} & \checkmark & Gene expression \\ \hline
MiNet & 2019 & ISBRA 2019 & Survival Analysis  & Reactome & \href{https://github.com/DataX-JieHao/MiNet}{\checkmark} & \checkmark & Gene expression, CNV, DNA methylation \\ \hline
pathDNN & 2020 & Journal of Chemical Information and Modeling & Drug sensitivity prediction & KEGG & \href{https://github.com/Charrick/drug_sensitivity_pred}{\checkmark} & \checkmark & Gene expression, drug targets \\ \hline
Multi-scale NN & 2020 & Plos one & Prediction of disease, pathway, and gene associations & Reactome & \href{https://life.bsc.es/iconbi/MultiScaleNN/index.html}{\checkmark} & \checkmark & Gene expression \\ \hline
GCN-MAE & 2020 & Bioinformatics & Cancer subtype classification  & KEGG & \href{https://github.com/smaster7/GCN_MAE}{Code not available} & X & Gene expression \\ \hline
P-NET & 2021 & Nature & Cancer state prediction & Reactome & \href{https://github.com/marakeby/pnet_prostate_paper}{\checkmark} & \checkmark & Mutations, CNA \\ \hline
PathCNN & 2021 & Bioinformatics & Binary classification Long-term VS Short-term survival & KEGG & \href{https://github.com/mskspi/PathCNN}{\checkmark} & NB: Only processed data & Gene expression, DNA methylation, CNV \\ \hline
PathDeep & 2021 & International Journal of Molecular Sciences & Classification cancer vs normal tissue & MSigDB & \href{https://github.com/sipark5340/PathDeep}{\checkmark} & NB: Only toy dataset available & Gene expression \\ \hline
PathGNN & 2022 & BMC Bioinformatics & Binary classification Long-term VS Short-term survival & Reactome & \href{https://github.com/BioAI-kits/PathGNN}{\checkmark} & \checkmark & Gene expression, clinical data \\ \hline
MPVNN & 2022 & Bioinformatics & Survival analysis & Unknown & \href{https://github.com/gourabghoshroy/MPVNN}{\checkmark} & \checkmark & Gene expression \\ \hline
GCS-Net & 2022 & Journal of Oncology & Binary classification Long-term VS Short-term survival & Reactome & Code not available & X & CNV, Somatic mutations, clinical data \\ \hline
ReGeNNe & 2023 & Bioinformatics & Classification (kidney stage, kidney vs liver, binary survival for ovarian) & PID, BioCarta, Reactome & \href{https://github.com/divya031090/ReGeNNe}{\checkmark} & X & Gene expression \\ \hline
BINN & 2023 & Nature Communications & Phenotypes classification & Reactome & \href{https://github.com/InfectionMedicineProteomics/BINN}{\checkmark} & \checkmark & Proteomic Data \\ \hline
PINNet & 2023 & Frontiers in Aging Neuroscience & Alzheimer disease classification & KEGG, GO BP & \href{https://github.com/DMCB-GIST/PINNet}{\checkmark} & \checkmark & Gene expression \\ \hline
PGLCN & 2023 & Computational and Structural Biotechnology Journal & Tumor mutation burden prediction  & Reactome & \href{https://github.com/liuchuwei/PGLCN}{X} & Github with empty files, not usable & Gene expression, CNV, Methylation \\ \hline
DeepKEGG & 2024 & Briefings in Bioinformatics & Cancer Recurrence Prediction - Binary classification  & KEGG & \href{https://github.com/lanbiolab/DeepKEGG}{\checkmark} & \checkmark & mRNA expression, SNV, miRNA \\ \hline
GraphPath & 2024 & Bioinformatics & Cancer status classification  & KEGG & \href{https://github.com/amazingma/GraphPath}{\checkmark} & \checkmark & CNA, Mutation \\ \hline
Pathformer & 2024 & Bioinformatics & Disease diagnosis and prognosis  & KEGG, PID, Reactome, BioCarta & \href{https://github.com/lulab/Pathformer}{\checkmark} & \checkmark & Gene expression (or multimodal) \\ \hline
Autosurv & 2024 & Precision Oncology & Survival Analysis  & Reactome & \href{https://github.com/jianglindong93/AUTOSurv}{\checkmark} & \checkmark & Gene expression, miRNA \\ \hline
\hline
\end{longtable}

 \begin{table}[]
	\centering
	\large
	\renewcommand{\arraystretch}{1.5}
	\resizebox{\textwidth}{!}{%
		\begin{tabular}{|>{\columncolor[HTML]{8FA98F}}c|c|c|c|c|}
			\hline
			\rowcolor[HTML]{D3AFAF}
			\textbf{Model} & \textbf{Metric} & \textbf{Pathway-informed model} & \textbf{Randomized model} & \textbf{Execution Time} \\
			\hline
			PASNet & AUC & 0.600 ± 0.067 & 0.608 ± 0.065 & ++ \\
			\hline
			CoxPASNet & C Index & 0.672 ± 0.002 & 0.672 ± 0.002 & ++ \\
			\hline
			MiNet & C Index & 0.650 ± 0.020 & 0.652 ± 0.025 & +++ \\
			\hline
			pathDNN & $R^{2}$ & 0.801 ± 0.007 & \textbf{0.806 ± 0.007} & +++ \\
			\hline
			MultiScaleNN & Accuracy & 0.660 ± 0.013 & 0.659 ± 0.014 & +++ \\
			\hline
			P-NET & AUC & 0.899 ± 0.021 & 
			\begin{tabular}{c}
				\textbf{OP} 0.896 ± 0.024 \\ 
				\textbf{PP} 0.887 ± 0.025 \\
				\textbf{OP + PP} 0.892 ± 0.027
			\end{tabular} & ++ \\
			\hline
			PathCNN & AUC & 0.745 ± 0.011 & 0.746 ± 0.007 & ++ \\
			\hline
			PathGNN & AUC & 0.693 ± 0.067 & 0.687 ± 0.060 & ++++ \\
			\hline
			MPVNN & C Index & 0.632 ± 0.081 & \textbf{0.645 ± 0.086} & +++ \\
			\hline
			BINN & Accuracy & 0.944 ± 0.023 & 
			\begin{tabular}{c}
				\textbf{OP} 0.958 ± 0.016 \\ 
				\textbf{OP + PP} 0.937 ± 0.017
			\end{tabular} & ++ \\
			\hline
			PINNet & AUC & 0.974 ± 0.062 & 0.974 ± 0.064 & + \\
			\hline
			DeepKEGG & AUC & 0.892 ± 0.088 & \textbf{0.897 ± 0.090} & ++ \\
			\hline
			Autosurv & C Index & 0.734 ± 0.048 & 0.732 ± 0.048 & +++ \\
			\hline
			GraphPath & Accuracy & 0.867 ± 0.026 & \textbf{PP} 0.878 ± 0.032 & ++++ \\
			\hline
			Pathformer & F1 Macro & 0.609 ± 0.077 & 
			\begin{tabular}{c}
				\textbf{OP} 0.614 ± 0.071 \\ 
				\textbf{OP + PP} 0.587 ± 0.067 
			\end{tabular}& ++++ \\
			\hline
		\end{tabular}%
	}
	\caption{
		\textbf{Table summarizing the performance comparison between \textit{pathway-informed} and \textit{randomized versions} of various deep learning models across different evaluation metrics.} Each model's performance is reported in terms of its specific metric (e.g., AUC, C-Index, Accuracy, R-squared), alongside the corresponding \textit{mean ± standard deviation} values. The table also includes the \textit{execution time} for each model, with a legend denoting the time required for 20 runs, categorized as follows: \textbf{+} represents seconds,  \textbf{++} represents minutes,
		\textbf{+++} represents hours, and \textbf{++++} represents days.
		For certain models, the performance is further divided into \textit{Omic-Pathway Network (OP)}, \textit{Pathway-Pathway Network (PP)}, or a combination of both (\textit{OP + PP}), to reflect the different configurations evaluated. \textbf{Bolded values} indicate cases where the randomized version outperformed the pathway-informed version. The results for the MPVNN and DeepKEGG models are the average outcomes across different tumor types considered (detailed for tumor type are in Additional Information, Table \ref{tab:deepkegg_comparison} and \ref{tab: MPVNN_comparison}).
	}
	\label{tableres}
\end{table}

\FloatBarrier
\newpage
\section*{Additional Information}
\FloatBarrier
\renewcommand{\thefigure}{A\arabic{figure}}
\renewcommand{\thetable}{A\arabic{table}}

\setcounter{figure}{0}
\setcounter{table}{0}

\subsection*{Mathematical Formalization of the Randomization Process}

Let $A$ be a binary matrix of dimensions $N \times M$:
\begin{equation}
    A = [a_{ij}], \quad \text{with } a_{ij} \in \{0,1\},
\end{equation}
where:
\begin{itemize}
    \item $N$ represents the number of features (e.g., genes),
    \item $M$ represents the number of pathways,
    \item $a_{ij} = 1$ indicates that feature $i$ is associated with pathway $j$, whereas $a_{ij} = 0$ indicates the absence of an association.
\end{itemize}

The total number of connections in the matrix is defined as:
\begin{equation}
    C = \sum_{i=1}^{N} \sum_{j=1}^{M} a_{ij}.
\end{equation}

Additionally, the number of features associated with each pathway $j$ is given by:
\begin{equation}
    d_j = \sum_{i=1}^{N} a_{ij}, \quad \forall j \in \{1, \dots, M\}.
\end{equation}

\subsubsection*{Randomization Constraints}
The randomization process consists of shuffling the connections $a_{ij}$ within the matrix while preserving the following constraints:

\begin{enumerate}
    \item \textbf{Preservation of the total number of connections:}
    \begin{equation}
        \sum_{i=1}^{N} \sum_{j=1}^{M} a'_{ij} = C,
    \end{equation}
    where $A' = [a'_{ij}]$ is the resulting matrix after randomization.

    \item \textbf{Preservation of the number of features per pathway:}
    \begin{equation}
        \sum_{i=1}^{N} a'_{ij} = d_j, \quad \forall j \in \{1, \dots, M\}.
    \end{equation}

    \item \textbf{Uniform sampling of connections:} The reassignment of connections is performed uniformly among all possible configurations satisfying the above constraints, ensuring that no structural bias or prior is introduced.\\
\end{enumerate}

\subsubsection*{Randomization Method}
The randomization operation can be performed through a uniform permutation of the connections while maintaining the above constraints. A possible algorithm for this process is:

\begin{enumerate}
    \item Extract a list of all existing $1$'s in matrix $A$ along with their respective indices $(i,j)$.
    \item Shuffle this list uniformly.
    \item Redistribute the $1$'s in the matrix $A'$ while ensuring that each column $j$ maintains the same number of connections $d_j$ as in the original matrix.
\end{enumerate}

This ensures that the matrix sparsity remains unchanged after the randomization process.\\

\subsubsection*{Randomization Trials}
In this trial, the randomization procedure described above was repeated 30 times, modifying the seed for the randomization functions, ensuring that each time a different set of connections was sampled among the $C$ possible ones. \\

\subsubsection*{Analysis of Optimal Sparsity Levels}
To analyze the optimal level of sparsity relative to the pathway-induced sparsity level, we define the total number of connections $C$ as the pathway-induced sparsity level. In a subsequent analysis, we allow the number of retained connections to vary between 60\% and 99\% of the total possible connections, i.e.,
\begin{equation}
    C' = k \cdot N M, \quad \text{con } k \in [0.6, 0.99]
\end{equation}

In this case, the randomization process is performed by uniformly sampling from all possible connections while ensuring that the total number of connections is equal to the desired sparsity level $C'$. However, the constraint on the number of connections per pathway is relaxed, allowing for a more flexible distribution of connections across pathways.\\

\subsection*{Comparison of Biological Information Extracted by Pathway-Informed Models and Randomized Counterparts}

For interpretability analysis, the methods reported in the original model papers were employed, otherwise permutation importance was used due to its simplicity. \\
For BINN, interpretability analyses were performed using SHAP (SHapley Additive exPlanations). Specifically, obtained SHAP values were adjusted using the logarithm of the number of nodes in each node’s reachable subgraph. This was done to take into account node connectivity and to avoid possible biases due to highly connected nodes. \\
SHAP methodology was also employed for PINNet. Precisely, the DeepExplainer implementation of SHAP, based on Deep SHAP, was used to calculate each input feature contribution to the model predictions. SHAP values were then aggregated across different cross-validation folds to obtain the attribution scores, which were then normalized via z-scores. \\
In the case of the DeepKEGG model, a simplified version of the DeepLIFT method was used. In this approach, the contribution of each feature to the model predictions is computed by multiplying the gradient of the output with respect to the input by the difference among the actual outputs and a reference activation (which was set to zero in this study). Feature importances were then obtained by aggregating across all samples to assess overall relevance. \\
In PASNet, since no interpretability module was provided in the GitHub repository of the model, a basic permutation importance approach was employed. Each feature was individually permuted, while keeping all the others fixed. Drops in performance were measured to estimate the features’ relevance to the model output. Finally, features were ranked in descending order of performance impact. \\
Regarding the PathCNN model, it was not possible to perform a fair comparison among biomarkers extracted by the pathway-informed model and its randomized counterpart. In fact, in the original paper, the interpretability analyses were carried out using Grad-CAM methods and focused on the pathway images provided as input to the model, thereby identifying entire pathways as important features. In such a setting, randomizing the pathway-related information fundamentally alters the input semantics, making any comparison of the most important features meaningless—since, in the randomized model, those features no longer correspond to actual biological pathways.

\subsection*{Graph Structure and Formal Definition of Metrics}

In our analysis, prior biological knowledge is encoded either as a bipartite graph, connecting features (e.g. genes) to pathways or as a simple graph, connecting pathways to one another. \\
The bipartite graph can be represented as \( G = (V_P \cup V_F, E) \), where \( V_P \) denotes the set of pathways and \( V_F \) the set of features. The cardinalities of the sets are \( |V_P| = M \) and \( |V_F| = N \), respectively. The set of edges \( E \subseteq V_P \times V_F \) represents known biological associations between pathways and features. By construction, this is a bipartite graph: edges only connect nodes of different types. \\
We denote the total number of nodes in the graph as \( M + N \), and the number of edges as \( |E| \). \\
The density of the bipartite graph is defined as the ratio between the number of observed edges and the number of possible edges. \\
In a complete bipartite graph with the same partition sizes, that is:

\begin{equation}
D = \frac{|E|}{M \cdot N}
\end{equation}

As the graphs of interest are typically sparse, we also report the sparsity level as \( S = 1 - D \). In Table \ref{tab:models_sparsity_table} it can be noted that the values of sparsity are rather high, ranging from 97.4 \% to 99.9\%.

The degree \( \deg(v) \) of a node \( v \in V_P \cup V_F \) corresponds to the number of adjacent edges. We define the average degree across all nodes as:

\begin{equation}
\bar{d} = \frac{2|E|}{M + N}
\end{equation}

Additionally, to capture structural asymmetries between the two partitions, we compute the average degree within each set separately: the average pathway degree is

\begin{equation}
\bar{d}_P = \frac{1}{M} \sum_{v \in V_P} \deg(v),
\end{equation}

while the average feature degree is

\begin{equation}
\bar{d}_F = \frac{1}{N} \sum_{v \in V_F} \deg(v).
\end{equation}

Where relevant, we also report median degrees, which can provide a more robust measure in the presence of hub nodes or heavy-tailed degree distributions.

Connectivity is further characterized by identifying the largest connected component \( C_{\text{max}} \subseteq V \). Its size, \( |C_{\text{max}}| \), indicates the number of nodes reachable from each other via paths in the graph. A large connected component suggests that most nodes belong to a single, globally connected subgraph, whereas multiple smaller components may reflect biological modularity or fragmentation.

The diameter of the graph, computed within the largest connected component, is defined as the greatest shortest-path distance between any two nodes in that component:

\begin{equation}
\text{diam}(G) = \max_{u,v \in C_{\text{max}}} d(u, v),
\end{equation}

where \( d(u, v) \) denotes the length of the shortest path between nodes \( u \) and \( v \).

Lastly, we evaluate degree assortativity, which measures the correlation between the degrees of connected nodes. \\
The assortativity coefficient is computed as:

\begin{equation}
r = \frac{
|E|^{-1} \sum_{(u,v) \in E} k_u k_v - \left[ |E|^{-1} \sum_{(u,v) \in E} \frac{1}{2}(k_u + k_v) \right]^2
}{
|E|^{-1} \sum_{(u,v) \in E} \frac{1}{2}(k_u^2 + k_v^2) - \left[ |E|^{-1} \sum_{(u,v) \in E} \frac{1}{2}(k_u + k_v) \right]^2
}
\end{equation}

where \( k_u \) and \( k_v \) denote the degrees of the nodes at the ends of each edge \( (u, v) \in E \), and \( |E| \) is the total number of edges in the graph. \\

 In our case, assortativity always assumes negative values, reflecting a disassortative structure in which highly connected nodes tend to link with low-degree nodes, and vice versa. \\
From a biological standpoint, this could be due to the fact that pathways often share common core genes (e.g., hub genes involved in multiple processes), resulting in a few features with very high degrees. On the other hand, many pathways include only a modest number of genes, creating a broad degree disparity. This enforces disassortative mixing, where hub features are connected to many low-degree pathways, lowering the assortativity coefficient. \\
Assortativity is also negative across all considered simple graphs connecting pathways to one another, as shown in Table \ref{tab:models_pathway_sparsity_table}. In that case, however, the absolute values of this parameter tend to be lower and closer to zero, meaning that the degrees of the nodes are likely to be more homogeneous. 

    \begin{table}[h]
        \centering
        \renewcommand{\arraystretch}{1.3}
        \begin{tabular}{|>{\columncolor[HTML]{8FA98F}}p{2.4cm}|p{2cm}|p{2cm}|p{2.5cm}|}
            \hline
            \rowcolor[HTML]{D3AFAF}
            \textbf{Model} & \textbf{Samples} & \textbf{Features} & \textbf{Sample to Feature Ratio} \\
            \hline
            PASNet & 464 & 4359 & 0.11 \\
            \hline
            CoxPASNet & 522 & 5567 & 0.094 \\
            \hline
            MiNet & 523 & 24803 & 0.021 \\
            \hline
            PathDNN & 198929 & 1278 & 160 \\
            \hline
            MultiScaleNN & 4788 & 9247 & 0.52 \\
            \hline
            P-NET & 1011 & 27687 & 0.037 \\
            \hline
            PathCNN & 287 & 4989 & 0.058 \\
            \hline
            PathGNN & 269 & 8611 & 0.031 \\
            \hline
            MPVNN & \multicolumn{3}{c|}{Variable (see Table \ref{tab:mpvnn_statistics})} \\
            \hline
            BINN & 197 & 554 & 0.36 \\
            \hline
            PINNet & 467 & 8922 & 0.052 \\
            \hline
            DeepKEGG & \multicolumn{3}{c|}{Variable (see Table \ref{tab:deepkegg_statistics})} \\
            \hline
            AutoSurv & 1058 & 3215 & 0.33 \\
            \hline
            GraphPath & 1013 & 12556 & 0.081 \\
            \hline
            Pathformer & 247 & 11560 & 0.021 \\
            \hline
        \end{tabular}
        \caption{\textbf{Overview of the analyzed models with their respective sample and feature statistics.}\\
        The Sample to Feature Ratio column represents the ratio of the number of samples to the number of features for each model, rounded to two significant figures. Models marked as "Variable" have different sample and feature sizes based on specific datasets (see referenced tables).}
        \label{tab:model_statistics}
    \end{table}

    \begin{table}[h]
        \centering
        \renewcommand{\arraystretch}{1.3}
        \begin{tabular}{|>{\columncolor[HTML]{8FA98F}}c|c|c|c|}
            \hline
            \rowcolor[HTML]{D3AFAF}
            \textbf{Tumor Type} & \textbf{Samples} & \textbf{Features} & \textbf{Sample to Feature Ratio} \\
            \hline
            BLCA & 426 & 1440 & 0.30 \\
            \hline
            BRCA & 1218 & 1440 & 0.85 \\
            \hline
            COADREAD & 434 & 1440 & 0.30 \\
            \hline
            GBM & 172 & 1440 & 0.12 \\
            \hline
            HNSC & 566 & 1440 & 0.39 \\
            \hline
            KIRC & 606 & 1440 & 0.42 \\
            \hline
            LIHC & 423 & 1440 & 0.29 \\
            \hline
            LUNG & 1129 & 1440 & 0.78 \\
            \hline
            OV & 308 & 1440 & 0.21 \\
            \hline
            STAD & 450 & 1440 & 0.31 \\
            \hline
        \end{tabular}
        \caption{\textbf{MPVNN: Tumor-specific sample and feature distributions.}\\
        The Sample to Feature Ratio column represents the ratio of the number of samples to the number of features for each tumor type, rounded to two significant figures.}
        \label{tab:mpvnn_statistics}
    \end{table}

    \begin{table}[h]
        \centering
        \renewcommand{\arraystretch}{1.3}
        \begin{tabular}{|>{\columncolor[HTML]{8FA98F}}c|c|c|c|}
            \hline
            \rowcolor[HTML]{D3AFAF}
            \textbf{Tumor Type} & \textbf{Samples} & \textbf{Features} & \textbf{Sample to Feature Ratio} \\
            \hline
            AML & 354 & 2200 & 0.16 \\
            \hline
            BLCA & 402 & 2100 & 0.19 \\
            \hline
            BRCA & 211 & 2100 & 0.10 \\
            \hline
            LIHC & 354 & 2200 & 0.16 \\
            \hline
            PRAD & 250 & 3600 & 0.069 \\
            \hline
            WT & 112 & 2200 & 0.051 \\
            \hline
        \end{tabular}
        \caption{\textbf{DeepKEGG: Tumor-specific sample and feature distributions.}\\
        The Sample to Feature Ratio column represents the ratio of the number of samples to the number of features for each tumor type, rounded to two significant figures.}
        \label{tab:deepkegg_statistics}
    \end{table}

  \begin{table}[h]
        \centering
        \renewcommand{\arraystretch}{1.2}
        \resizebox{\textwidth}{!}{ 
        \begin{tabular}{|>{\columncolor[HTML]{8FA98F}}l|c|c|c|c|c|c|c|c|c|c|}
            \hline
            \rowcolor[HTML]{D3AFAF}
            \textbf{Model} & \makecell{Number \\ of \\ Pathways} & \makecell{Number \\ of \\ Nodes} & \makecell{Number \\ of \\ Edges} & \makecell{Density / \\ Sparsity Level} & \makecell{Average \\ Degree} & \makecell{Largest \\ Component \\ Size} & \makecell{Diameter} & \makecell{Assortativity} & \makecell{Average / Median \\ Degree \\ (Pathways)} & \makecell{Average / Median \\ Degree \\ (Features)} \\
            \hline
            PASNet & 574 & 4934 & 28171 & 0.011 / 98.9\% & 11.4 & 4897 & - & -0.26 & 49.1 / 26 & 6.5 / 4 \\
            \hline
            CoxPASNet & 860 & 6428 & 39609 & 0.008 / 99.2\% & 12.3 & 6427 & - & -0.25 & 46.1 / 25 & 7.1 / 5 \\
            \hline
            MiNet & 507 & 5481 & 34955 & 0.003 / 99.7\% & 12.8 & 5481 & 6 & -0.29 & 75.0 / 54 & 12.8 / 4 \\
            \hline
            PathDNN & 323 & 1600 & 10695 & 0.026 / 97.4\% & 13.4 & 1587 & - & -0.2 & 33.2 / 19 & 8.4 / 3 \\
            \hline
            MultiScaleNN & 1708 & 10805 & 20904 & 0.001 / 99.9\% & 3.9 & 10805 & 11 & -0.13 & 3.9 / 1 & 14.1 / 9 \\
            \hline
            P-NET & 2029 & 10690 & 103351 & 0.002 / 99.8\% & 19.3 & 10690 & 5 & -0.19 & 60.4 / 28 & 19.3 / 6 \\
            \hline
            PathCNN & 146 & 4969 & 9905 & 0.014 / 98.6\% & 4.0 & 4969 & 8 & -0.46 & 69.9 / 51 & 4.0 / 1 \\
            \hline
            PathGNN & \multicolumn{10}{c|}{Not Applicable}  \\
            \hline
            MPVNN & 1 & 354 & 3092 & 0.025 / 97.5\% & 17.5 & 322 & - & -0.39 & - & - \\
            \hline
            BINN & 2585 & 11613 & 45820 & 0.032 / 96.8\% & 7.9 & 11613 & 10 & -0.2 & 26.3 / 15 & 7.9 / 2 \\
            \hline
            PINNet & 168 & 9090 & 7095 & 0.005 / 99.5\% & 1.6 & 2753 & - & -0.45 & 42.2 / 33 & 0.8 / 1 \\
            \hline
            DeepKEGG & \multicolumn{10}{c|}{Variable (see Table \ref{tab:deepkegg})}  \\
            \hline
            Autosurv & 581 & 7302 & 22890 & 0.012 / 98.8\% & 6.3 & 7174 & - & -0.35 & 39.4 / 28 & 3.4 / 2 \\
            \hline
            GraphPath  & \multicolumn{10}{c|}{Not Applicable}  \\
            \hline
            Pathformer & 1497 & 11560 & 86460 & 0.005 / 99.5\% & 15.0 & 11560 & 6 & -0.15 & 64.9 / 44 & 15.0 / 5 \\
            \hline
        \end{tabular}
        } 
        \caption{\textbf{Pathway-Feature Network properties of different models.}}
        \label{tab:models_sparsity_table}
    \end{table}

\begin{table}[h]
    \centering
    \renewcommand{\arraystretch}{1.2}
    \begin{tabular}{|>{\columncolor[HTML]{8FA98F}}l|c|c|c|}
        \hline
        \rowcolor[HTML]{D3AFAF}
        \textbf{Tumor} & \makecell{\textbf{SNV-Pathway} \\ \textbf{Sparsity Level}} & \makecell{\textbf{mRNA-Pathway} \\ \textbf{Sparsity Level}} & \makecell{\textbf{miRNA-Pathway} \\ \textbf{Sparsity Level}} \\
        \hline
        AML  & 98.6\% & 98.8\% & 63.7\% \\
        \hline
        BLCA & 98.8\% & 98.8\% & 66.9\% \\
        \hline
        BRCA & 98.7\% & 98.7\% & 63.3\% \\
        \hline
        LIHC & 98.6\% & 98.8\% & 63.7\% \\
        \hline
        PRAD & 98.7\% & 98.7\% & 62.7\% \\
        \hline
        WT   & -    & 98.7\% & 59.2\% \\
        \hline
    \end{tabular}
    \caption{\textbf{Feature-Pathway Sparsity Levels for DeepKEGG across different tumor types.}}
    \label{tab:deepkegg}
\end{table}

    \begin{table}[h]
        \centering
        \renewcommand{\arraystretch}{1.2}
        \resizebox{\textwidth}{!}{ 
        \begin{tabular}{|>{\columncolor[HTML]{8FA98F}}l|c|c|c|c|c|c|c|c|}
            \hline
            \rowcolor[HTML]{D3AFAF}
            \textbf{Model} & \makecell{Number \\ of \\ Pathways} & \makecell{Number \\ of \\ Nodes} & \makecell{Number \\ of \\ Edges} & \makecell{Density / \\ Sparsity Level} & \makecell{Average \\ Degree} & \makecell{Largest \\ Component \\ Size} & \makecell{Diameter} & \makecell{Assortativity} \\
            \hline
            P-NET & 23441 & 23441 & 23659 & 0.00 / 99.99\% & 2.02 & 1540 & - & -0.17 \\
            \hline
            BINN & 2585 & 2585 & 2603 & 0.00 / 99.92 \% & 2.01 & 1040 & - & -0.157  \\
            \hline
            GraphPath & 511 & 511 & 2245 & 0.01 / 99.14 \% & 8.79 & 429 & - & -0.05  \\
            \hline
            Pathformer & 1497 & 1497 & 222693 & 0.10 / 90.06\% & 297.52 & 1326 & - & -0.01 \\
            \hline
        \end{tabular}
        } 
        \caption{\textbf{Pathway-Pathway Network properties of different models.}}
        \label{tab:models_pathway_sparsity_table}
    \end{table}

\begin{table}[h]
\centering
\begin{tabular}{|>{\columncolor[HTML]{8FA98F}}c|c|c|}
\hline
\rowcolor[HTML]{D3AFAF}
\hline
\textbf{Tumor Type} & \textbf{DeepKEGG - Pathway-informed} & \textbf{DeepKEGG - Randomized } \\
\hline
AML & 0.960 ± 0.022 & 0.966 ± 0.017 \\
\hline
BLCA & 0.951 ± 0.020 & \textbf{0.973 ± 0.013} \\
\hline
BRCA & 0.868 ± 0.059 & 0.860 ± 0.044 \\
\hline
LIHC & 0.959 ± 0.022 & 0.963 ± 0.017 \\
\hline
PRAD & 0.777 ± 0.056 & 0.776 ± 0.059 \\
\hline
WT   & 0.839 ± 0.094 & 0.844 ± 0.091 \\
\hline

\end{tabular}
\caption{\textbf{Comparison of AUC values for DeepKEGG model across different tumor types.}
Additional trials for the DeepKEGG model were conducted on the LIHC tumor type, which, along with BLCA, was highlighted in the original paper as a case study for biomarker discovery. We selected LIHC over BLCA to ensure more comparable performance between pathway-informed models and their randomized counterparts.}
\label{tab:deepkegg_comparison}
\end{table}

\begin{table}[h]
\centering
\begin{tabular}{|>{\columncolor[HTML]{8FA98F}}c|c|c|}
\hline
\rowcolor[HTML]{D3AFAF}
\hline
\textbf{Tumor Type} & \textbf{MPVNN - Pathway-informed } & \textbf{MPVNN - Randomized } \\
\hline
BLCA & 0.689 ± 0.019 & \textbf{0.701 ± 0.01} \\
\hline
BRCA & 0.723 ± 0.011 & \textbf{0.756 ± 0.017} \\
\hline
COADREAD & 0.651 ± 0.013 & \textbf{0.746 ± 0.042} \\
\hline
GBM & 0.600 ± 0.013 & \textbf{0.636 ± 0.029} \\
\hline
HNSC & 0.537 ± 0.011 & \textbf{0.585 ± 0.012} \\
\hline
KIRC & 0.740 ± 0.010 & 0.723 ± 0.061 \\
\hline
LIHC & \textbf{0.710 ± 0.008} & 0.598 ± 0.019 \\
\hline
LUNG & 0.619 ± 0.004 & 0.619 ± 0.002 \\
\hline
OV & \textbf{0.504 ± 0.025} & 0.480 ± 0.005 \\
\hline
STAD & 0.546 ± 0.026 & \textbf{0.609 ± 0.015} \\
\hline
\end{tabular}
\caption{\textbf{Comparison of C-Index values for MPVNN model across different tumour types.} The results of the MPVNN model are characterized by a notable imbalance in performance between the pathway-informed model and its randomized counterparts. Despite the overall performance being significantly higher in the randomized version of the model, there are few tumour types where the pathway-informed version of MPVNN outperforms its randomized equivalent. This greater variability could be explained by the fact that the architecture of the MPVNN model relies  on a small sets of genes connected by signal flow within the PI3K-Akt pathway. Given that the model is built on a single specific pathway, any perturbations can lead to pronounced effects, often skewing performance towards the randomization. However, there are few instances for which the perturbation may also enhance the performance of the pathway-informed version.}
\label{tab: MPVNN_comparison}
\end{table}

 \begin{figure}[ht]
 \centering
 \includegraphics[width=\linewidth]{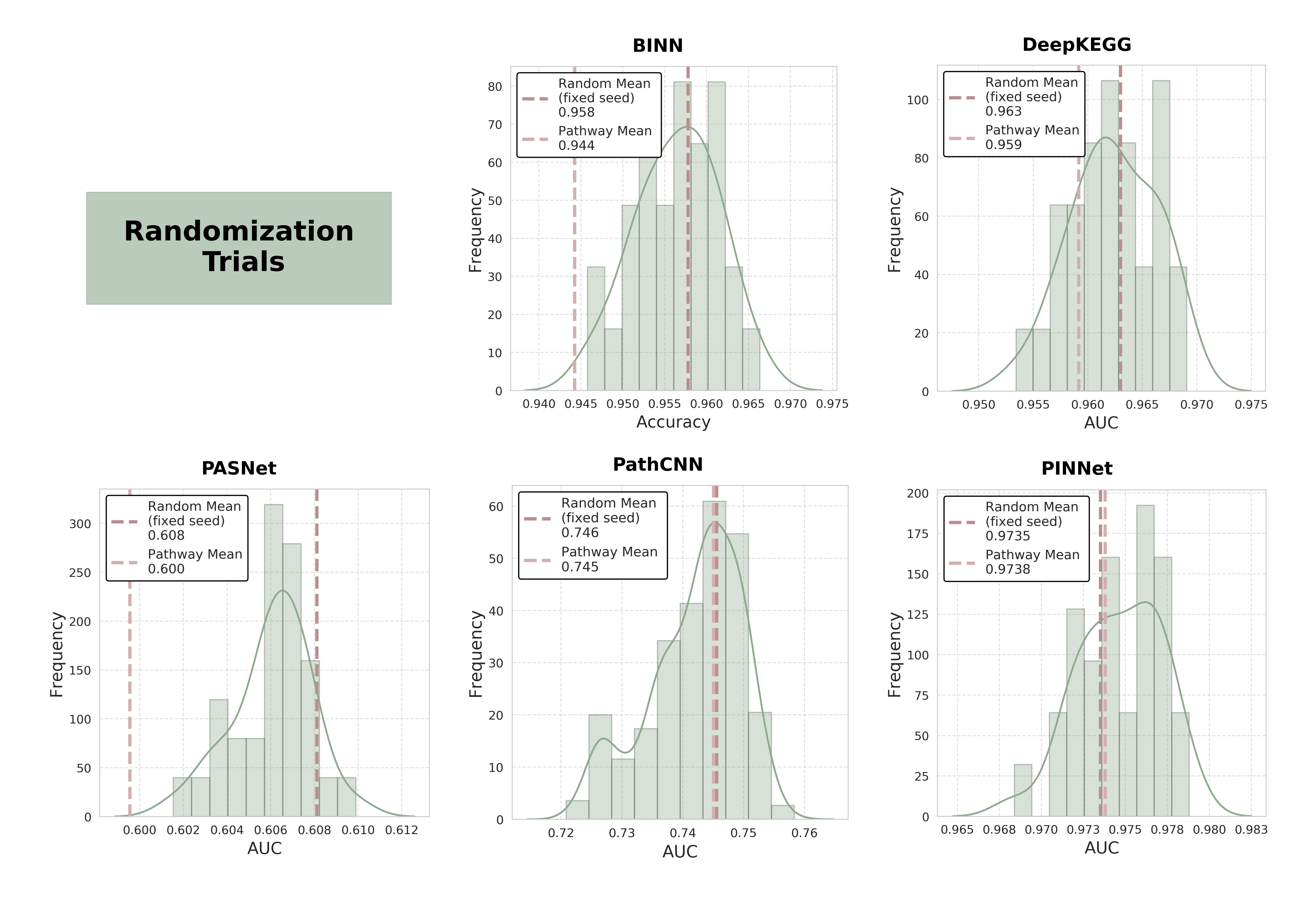}
 \caption{\textbf{Performance distribution across randomized trials for BINN, DeepKEGG, PASNet, PathCNN, and PINNet}.
Histograms represent the distribution of model performance (AUC or Accuracy) obtained by randomizing pathway-related information using different random seeds. The dashed lines indicate the mean performance obtained with a fixed random seed, both for the pathway-informed model and its randomized counterpart. The results demonstrate that the fixed-seed performance aligns with the broader distribution of randomized trials, confirming that model outcomes are not biased by a particularly favorable random seed.}
 \label{fig:rand_trials}
 \end{figure}

  \begin{figure}[ht]
 \centering
 \includegraphics[width=\linewidth]{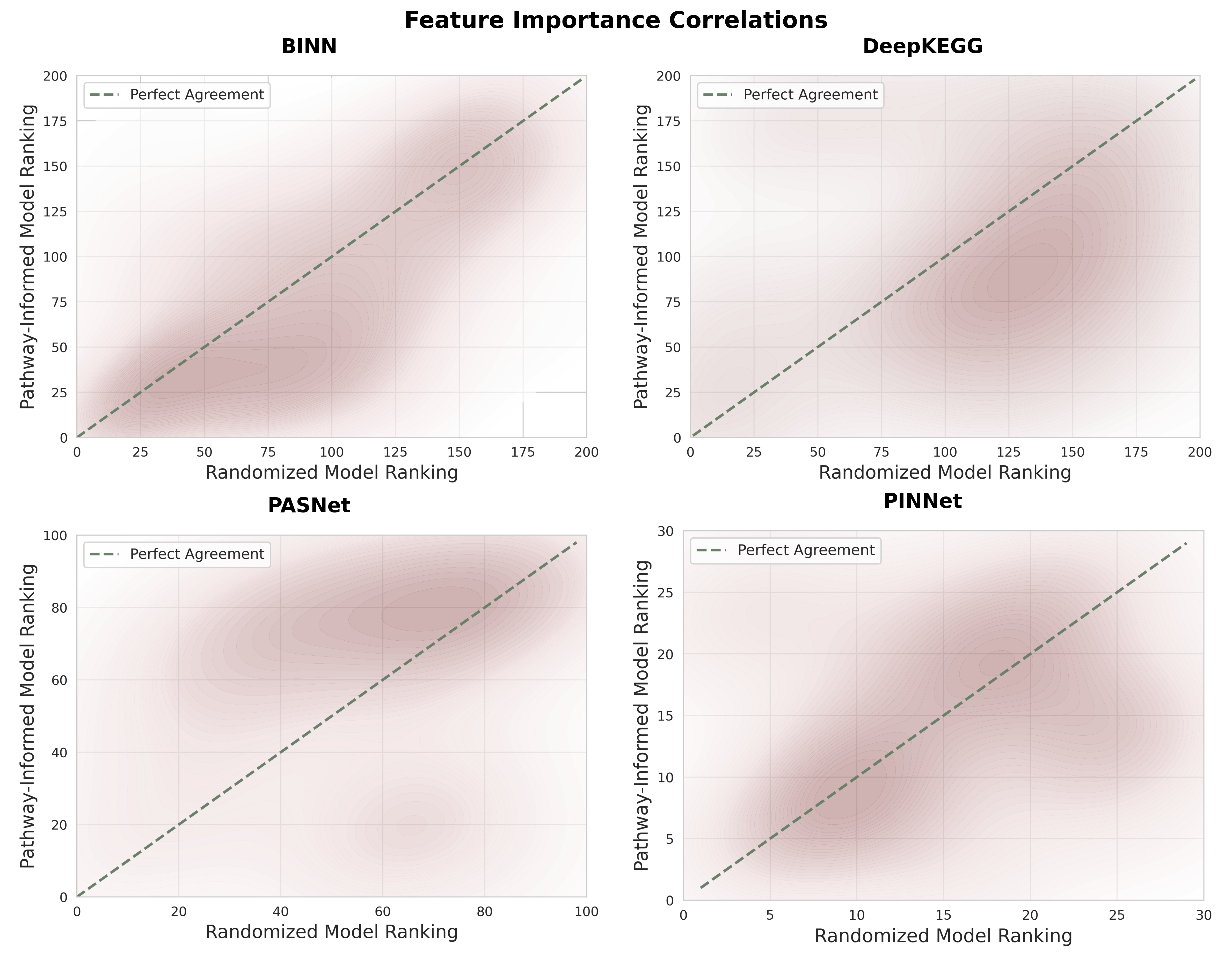}
 \caption{\textbf{ Comparison of feature importance rankings between pathway-informed and randomized models across BINN, DeepKEGG, PASNet, and PINNet.}.
 Each scatter density plot illustrates the relationship between feature rankings in the pathway-informed and randomized versions of the respective models. The x-axis represents feature rankings in the randomized model, while the y-axis represents rankings in the pathway-informed model. The dashed diagonal line indicates perfect agreement between the two ranking sets. The density contours highlight the concentration of ranked features, showing the extent of alignment or deviation between the models.}
 \label{fig:importance_trials}
 \end{figure}

\begin{figure}[ht]
	\centering
	\includegraphics[width=\linewidth]{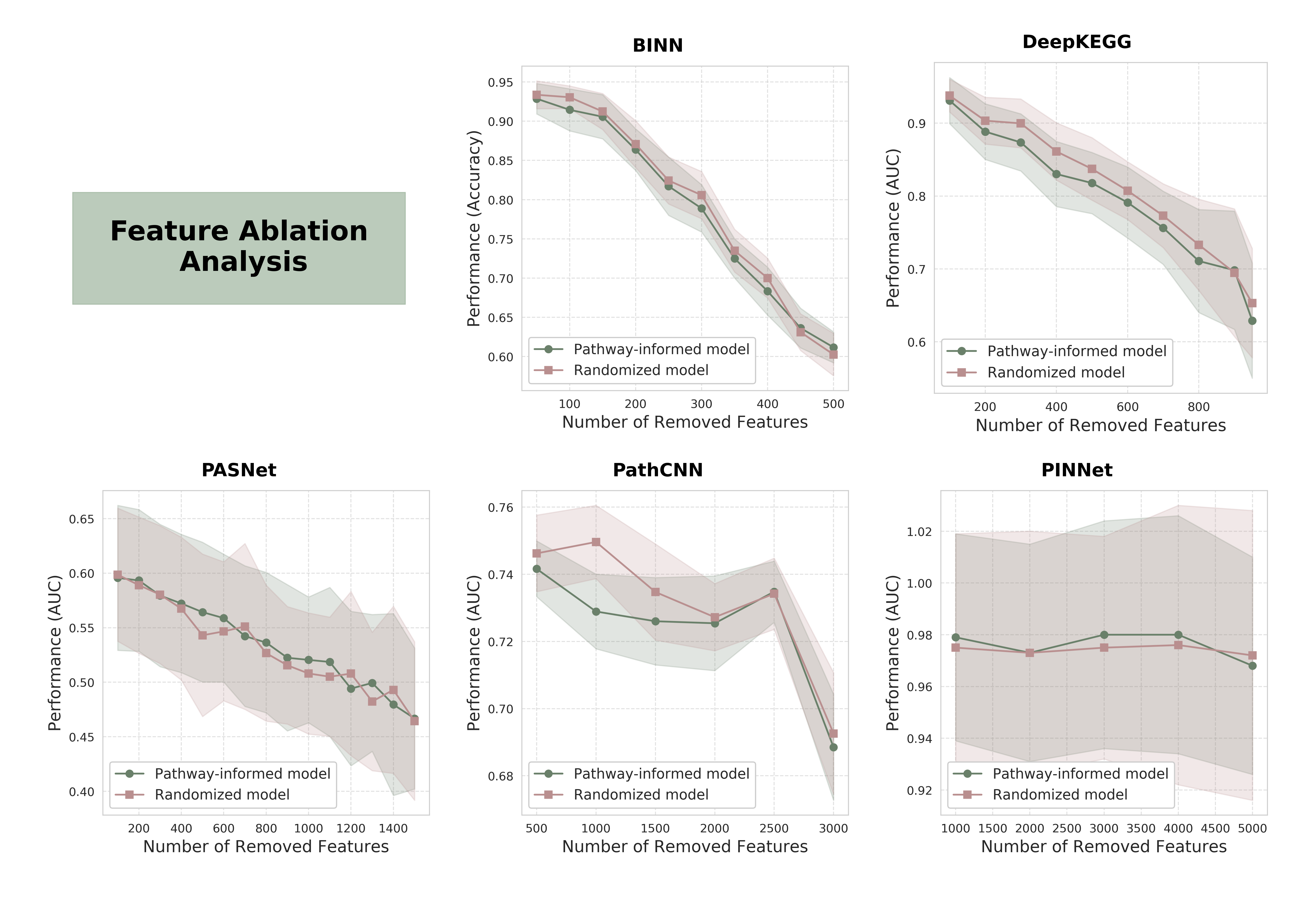}
	\caption{ \textbf{Impact of feature ablation on model performance}.
    Feature Ablation Analysis: Performance degradation as an increasing number of highly discriminative features are removed, comparing pathway-informed models (green) with their randomized counterparts (pink). Shaded areas represent the standard deviation across runs. Across all models, performance declines with feature removal, and pathway-informed models do not consistently outperform randomized models.}
	\label{fig:main_ablation}
	\end{figure}

\subsection*{Data and Code availability}

The codes and datasets used to train the models were obtained from their respective repositories. Below is a list of the models along with the links to their repositories.

\begin{itemize}
    \item \textbf{PASNet} (2018, June 2024 version) - \href{https://github.com/DataX-JieHao/PASNet}{https://github.com/DataX-JieHao/PASNet}
    
    \item \textbf{CoxPASNet} (2018, June 2024 version) - \href{https://github.com/DataX-JieHao/Cox-PASNet}{https://github.com/DataX-JieHao/Cox-PASNet}
    
    \item \textbf{MiNet} (2019, June 2024 version) - \href{https://github.com/DataX-JieHao/MiNet}{https://github.com/DataX-JieHao/MiNet}
    
    \item \textbf{pathDNN} (2020, June 2024 version) - \href{https://github.com/Charrick/drug_sensitivity_pred}{https://github.com/Charrick/drug\_sensitivity\_pred}
    
    \item \textbf{Multi-scale NN} (2020, June 2024 version) - \href{https://life.bsc.es/iconbi/MultiScaleNN/index.html}{https://life.bsc.es/iconbi/MultiScaleNN/index.html}
       
    \item \textbf{PathCNN} (2021, June 2024 version) - \href{https://github.com/mskspi/PathCNN}{https://github.com/mskspi/PathCNN}
    
    \item \textbf{P-NET} (2021, September 2024 version) - \href{https://github.com/marakeby/pnet_prostate_paper}{https://github.com/marakeby/pnet\_prostate\_paper}
     
    \item \textbf{PathGNN} (2022, June 2024 version) - \href{https://github.com/BioAI-kits/PathGNN}{https://github.com/BioAI-kits/PathGNN}
    
    \item \textbf{MPVNN} (2022, June 2024 version) - \href{https://github.com/gourabghoshroy/MPVNN}{https://github.com/gourabghoshroy/MPVNN}    
    
    \item \textbf{BINN} (2023, June 2024 version) - \href{https://github.com/InfectionMedicineProteomics/BINN}{https://github.com/InfectionMedicineProteomics/BINN}

    \item \textbf{PINNet} (2023, June 2024 version) - \href{https://github.com/DMCB-GIST/PINNet}{https://github.com/DMCB-GIST/PINNet}
    
    \item \textbf{DeepKEGG} (2024, June 2024 version) - \href{https://github.com/lanbiolab/DeepKEGG}{https://github.com/lanbiolab/DeepKEGG}
    
    \item \textbf{GraphPath} (2024, July 2024 version) - \href{https://github.com/amazingma/GraphPath}{https://github.com/amazingma/GraphPath}
    
    \item \textbf{Autosurv} (2024, July 2024 version) - \href{https://github.com/jianglindong93/AUTOSurv}{https://github.com/jianglindong93/AUTOSurv}

    \item \textbf{Pathformer} (2024, July 2024 version) - \href{https://github.com/lulab/Pathformer}{https://github.com/lulab/Pathformer}
\end{itemize}

\end{document}